\documentclass[twocolumn]{aastex631}

\newcommand{\pr}{\ensuremath{P_{\rm r}}}

\newcommand{\pq}{\ensuremath{P_Q}}
\newcommand{\pu}{\ensuremath{P_U}}

\newcommand{\nnq}{\ensuremath{N_Q}}
\newcommand{\nnu}{\ensuremath{N_U}}
\newcommand{\ainv}{\ensuremath{\alpha_{0}} }

\newcommand{\pmax}{\ensuremath{P_{\rm{max}}}}

\usepackage{amsmath}
\usepackage{wasysym}

\begin{document}
\title{Polarimetric Behaviour of (3200) Phaethon: New Observations with Implications for Surface Heterogeneity}

\author[0000-0002-6610-1897]{Zuri Gray}
\affiliation{Department of Physics, PO Box 64, FI-00014 University of Helsinki, Finland}

\author[0009-0007-9187-6460]{Mikko Vuori}
\affiliation{Department of Physics, PO Box 64, FI-00014 University of Helsinki, Finland}

\author[0000-0002-9870-123X]{Eric MacLennan}
\affiliation{Department of Physics, PO Box 64, FI-00014 University of Helsinki, Finland}

\author[0000-0002-5624-1888]{Mikael Granvik}
\affiliation{Department of Physics, PO Box 64, FI-00014 University of Helsinki, Finland}
\affiliation{Asteroid Engineering Laboratory, Lule\r{a} University of Technology, Box 848, SE-98128 Kiruna, Sweden}

\author[0000-0001-8058-2642]{Karri Muinonen}
\affiliation{Department of Physics, PO Box 64, FI-00014 University of Helsinki, Finland}

\author[0000-0001-7403-1721]{Antti Penttil\"a}
\affiliation{Department of Physics, PO Box 64, FI-00014 University of Helsinki, Finland}

\author[0009-0009-1581-1408]{Mikael Turkki}
\affiliation{Finnish Centre for Astronomy with ESO (FINCA), FI-20014, University of Turku, Finland}
\affiliation{Department of Physics and Astronomy,University of Turku, FI-20014, University of Turku, Finland}
\affiliation{Aalto University Metsähovi Radio Observatory, Metsähovintie 114, 02540 Kylmälä, Finland}

% \keywords{}

\begin{abstract}

The near-Earth asteroid (3200) Phaethon exhibits unusually high linear polarisation at large phase angles.
Previous observations revealed a significant difference between apparitions: measurements obtained in 2016 showed a polarisation up to $\sim10$ percentage points higher than those obtained in 2017 at comparable phase angles. The distinct surface regions and viewing geometries sampled during the two apparitions have been suggested as possible explanations for this difference. Here, we present new R-band polarimetric observations obtained with the Nordic Optical Telescope, covering phase angles $32^\circ<\alpha<103^\circ$. The 2019 measurements show polarisation intermediate between higher and lower values measured in 2016 and 2017, respectively. Using a recent shape model, we calculate the distribution of reflected flux across the surface during the three individual apparitions and find that the 2019 observations sample the surface more similarly to 2016 than to 2017. We explore several possible explanations for the different polarisation behaviours, including differences in regolith grain size, related surface albedo, and observing geometry. While the available observations do not allow these effects to be uniquely disentangled, our results and analysis suggest that surface heterogeneity is the most plausible underlying cause, with viewing geometry potentially contributing to the observed polarimetric differences. 

\end{abstract}

% ===================================================================================================

\section{Introduction}
\label{intro}
The near-Earth asteroids (3200) Phaethon ($D\sim6\,$km) and (155140) 2005~UD ($D\sim1\,$km), and the Geminid meteor stream are thought to belong to a common dynamical complex produced by past disruption events \citep{ohtsuka2006,jewitt2006,kinoshita2007}. This connection is remarkable because it links asteroids, meteoroids, and a meteor shower, providing a rare opportunity to study how small bodies are disrupted, thermally altered, and transported through the inner Solar System \citep{granvik2016}. Phaethon is particularly intriguing because, despite its asteroid classification, it has been suspected to exhibit comet-like activity near perihelion and is the target of JAXA's upcoming \textit{DESTINY$^{+}$} mission. The inferred mass-loss rate of $\sim3\,$kg\,s$^{-1}$ is insufficient to sustain the present-day Geminid stream \citep{jewitt2013}, suggesting that past activity was much stronger or other disruption mechanisms were involved. One proposed mechanism is thermally driven decomposition of surface minerals during Phaethon's extreme perihelion heating \citep{maclennan2024}, resulting in gas and dust release, as well as spatially variable thermal alteration due to non-uniform heating. 

Polarimetry is a powerful remote-sensing technique that is particularly sensitive to the physical properties of airless surfaces, including regolith grain size, porosity, roughness, and composition. Phaethon has been observed polarimetrically during several apparitions \citep{ito2018,shinnaka2018,devogele2018,okazaki2020,kiselev2022,geem2022}, revealing exceptionally high degrees of linear polarisation of $40-50\%$ at large phase angles. However, these observations revealed an unexpected difference. The 2016 observations of \citet{ito2018} show substantially higher polarisation than the 2017 observations of \citet{shinnaka2018} and \citet{devogele2018}, with differences reaching $\sim10$ percentage points at the largest observed phase angles. The viewing geometry of Phaethon varied substantially between these apparitions. \citet{shinnaka2018} and \citet{kiselev2022} suggested that the difference could, therefore, be related to the different surface regions sampled, potentially indicating latitude-dependent scattering properties. The origin of this unusual polarimetric behaviour remains unresolved. 

Evidence for surface heterogeneity has also emerged from a variety of techniques, including spectroscopic observations \citep{kareta2018, lazzarin2019, lee2019, ohtsuka2020, kiselev2022, Betzler2024, Angrisani2025}, radar observations \citep{taylor2019}, geophysical modelling \citep{jo2025}, and thermal inertia modelling \citep{maclennan2022}. Several authors suggest that the heterogeneity of regolith properties is latitudinal \citep{lazzarin2019, lee2019, maclennan2022, Betzler2024, Angrisani2025}, while others suggest it is longitudinal \citep{taylor2019,ohtsuka2020,jo2025}. Together, these studies suggest that different regions of Phaethon's surface may possess distinct physical properties, although the origin, extent, and distribution of these variations remain uncertain.

In this work, we present new polarimetric observations of Phaethon obtained during its 2019 apparition, providing an important new viewing geometry compared to those sampled during the 2016 and 2017 campaigns. In addition, we calculate the distribution of reflected flux across Phaethon's surface across the three apparitions using a recent shape model and explore possible explanations for the measured differences in polarimetric behaviour. 

% ===================================================================================================
\section{Observations and Analysis}

\subsection{Polarimetric Observations}
We obtained imaging polarimetric observations of Phaethon between July and October, 2019, using the Nordic Optical Telescope (NOT) in the R-band (filter \#76, R\_Bes 650\_130), covering phase angles $32^{\circ}<\alpha<103^{\circ}.$ The polarimetric setup consists of a rotatable ($\lambda/2$) retarder waveplate mounted in the FAPOL unit, followed by a calcite beam-splitter. The fluxes of the ordinary and extraordinary beams in the resultant images were used to measure the polarisation of Phaethon. Using non-sidereal tracking, the observing strategy consisted of repeated polarimetric cycles, each comprising four individual 60-s exposures with the waveplate at position angles of 0, 22.5, 45, and 67.5$^{\circ}$. 

During the first few observing nights, Phaethon was observable only at low elevation ($\leq 20-30^{\circ}$) for less than 1~hr before sunrise, allowing only 4--6 polarimetric cycles per night. At later epochs, we typically obtained 12--13 polarimetric cycles per night, observing continuously for about 1.25~hr. On August 31 and September 9, we obtained extended time series of 36--37 cycles over $\sim$4-hr to investigate possible variations in polarisation over Phaethon's 3.6-hr rotation period \citep{hanus2016}.

To ensure the quality and reliability of the measurements, each exposure was inspected by eye. Frames in which the target photocentre in either the ordinary or extraordinary beam was contaminated by a background star were discarded, and as a consequence, its polarimetric pair. Unfortunately, the dense stellar field surrounding Phaethon resulted in substantial frame rejection, particularly at the earliest epochs.

Following \citet{bagnulo2009}, we derived the reduced Stokes parameters (\pq = $Q/I$, \pu = $U/I$) using the beam-swapping technique and calculated the angle of polarisation relative to the direction perpendicular to the scattering plane as $\theta = 0.5 \arctan(\pu/\pq)$. Since all observations were at phase angles $\alpha>20^{\circ}$, where Phaethon's polarisation is expected to be positive, the polarisation vector is expected to be aligned perpendicular to the scattering plane, and hence, $\theta$ and \pu~should be consistent with zero within the measurement uncertainties. As an additional quality check, we calculated the so-called Null Stokes parameters, which provide an estimate of the level of systematic errors by measuring differences between the Stokes parameters obtained from individual polarimetric pairs. 

In addition to Phaethon, we observed zero- and high-polarisation standard stars to measure and account for effects of instrumental polarisation. Standard bias and domeflat corrections were applied during pre-processing.

% ===================================================================================================
\subsection{Modelling the surface-flux distribution}

In this study, we compare high-phase-angle polarimetric measurements obtained during the 2016, 2017, and 2019 apparitions. The viewing geometry of Phaethon varied substantially between these apparitions, resulting in different portions of its surface being illuminated and visible to the observer. 

To investigate the surface regions contributing to the observations, we use the latest shape model to model the illumination and viewing geometry of Phaethon at each observing epoch. This Phaethon model has a spin pole in ecliptic coordinates of $\lambda_{eclip} = 318.1^{\circ}$, $\beta_{eclip} = -52.5^{\circ}$ and is derived from radar, lightcurve, and occultation observations \citep{marshall_acm2026}. We take an approach similar to \citet{maclennan2022}, wherein the latitude-dependent thermal flux contribution was calculated for a shape model\footnote{The shape model used in this work has similar spin pole coordinates, but notably differs from the 2022 model in that it is shorter in the z-axis dimension by around 16\% due to constraints from recent occultation observations.} of Phaethon. From the known positions of Phaethon, the Sun, and Earth at each observing epoch, we determine the sub-solar and sub-observer latitudes, $\beta_{\odot}$ and $\beta_{\oplus}$, respectively. We then calculate the relative contribution of the reflected sunlight across the surface by taking the cosine of the incident ($\mu_0$) and emergent ($\mu$) angles. We model Lommel-Seeliger scattering behaviour across the surface \citep{lumme1981} by computing the quantity, $\mu_0 \mu / (\mu_0 + \mu)$, for each shape facet then taking its average over a full rotation. For this calculation the light scattering properties of the facets are taken to be equal. After normalizing to the disk-integrated value, this approach approximates the relative contribution of each facet to the total reflected sunlight received by the observer. 

In lieu of the actual latitude, which gives the position of a facet relative to the centre of the shape, the relative contributions are grouped by facet tilt, which is defined as the orientation of the facet relative to the direction perpendicular to the spin pole.
Phaethon's top-spin shape is composed of large portions of surface areas that have similar surface tilt at the mid-latitudes and thus contribute to the observed flux in a similar manner.
Facet tilt broadly correlates with geographic latitude, with positive and negative facet tilts approximately corresponding to the northern and southern surface regions.

Figures~\ref{fig:ito}, \ref{fig:shinnaka}, and \ref{fig:not} show the resulting flux distributions for the 2016 \citep{ito2018}, 2017 \citep{shinnaka2018}, and 2019 observations, respectively. The vertical axes represent the tilt of the surface facets relative to the equator, which has a tilt of zero, while the horizontal axes give the relative contribution to the total reflected sunlight using a bin width of 6$^{\circ}$. Thus, the blue filled distribution illustrate how the disk-integrated reflected flux is distributed across the facet tilts in each epoch. 

These distributions allow us to compare the surface regions sampled during the three apparitions and investigate whether differences in polarisation correlate with the viewing and illumination geometry.

\begin{table*}
\caption{\label{tablePQ} Results of aperture polarimetric measurements of Phaethon, nightly average. $^{\rm{a}}$Heliocentric distance; $^{\rm{b}}$geocentric distance; $^{\rm{c}}$number of polarimetric pairs used for measurement; $^{\rm{d}}$phase angle; $^{\rm{e,f}}$reduced Stokes parameters; $^{\rm{g,h}}$null Stokes parameters; $^{\rm{i}}$polarisation angle}
\centering
\begin{tabular}{ccccccr@{$\pm$}lr@{$\pm$}lccr@{$\pm$}l}
\hline \hline

Date & UT & $^{\rm{a}}$r (au) & $^{\rm{b}}$$\Delta$ (au) & $^{\rm{c}}$N & $^{\rm{d}}$$\alpha~(^\circ)$ & \multicolumn{2}{c}{$^{\rm{e}}$\pq (\%)} & \multicolumn{2}{c}{$^{\rm{f}}$\pu (\%)} & $^{\rm{g}}$\nnq & $^{\rm{h}}$\nnu & \multicolumn{2}{c}{$^{\rm{i}}$$\theta~(^\circ)$} \\

\hline

2019-Jul-14 & 05:05 & 0.42 & 0.84 & 4 & 102.55 & 40.66 & 1.17 & -1.03 & 1.17 & $-$ & $-$ & 179.27 & 0.82 \\
2019-Jul-15 & 05:19 & 0.45 & 0.84 & 8 & 98.81 & 38.66 & 2.3 & -0.86 & 2.21 & -0.15 & -1.56 & 179.36 & 1.63 \\
2019-Jul-16 & 05:16 & 0.48 & 0.85 & 8 & 95.51 & 38.29 & 0.44 & -0.4 & 0.45 & -0.03 & 0.17 & 179.7 & 0.33 \\
2019-Jul-18 & 05:23 & 0.53 & 0.87 & 8 & 89.86 & 33.51 & 1.28 & 0.04 & 1.3 & 0.01 & 0.4 & 0.04 & 1.11  \\
2019-Jul-25 & 04:56 & 0.69 & 0.93 & 24 & 76.51 & 27.78 & 0.15 & -0.11 & 0.15 & 0.1 & -0.03 & 179.89 & 0.16  \\
2019-Jul-26 & 05:03 & 0.71 & 0.93 & 48 & 75.11 & 25.68 & 0.14 & -0.28 & 0.14 & -0.04 & -0.01 & 179.69 & 0.16  \\
2019-Jul-30 & 04:43 & 0.79 & 0.96 & 24 & 70.37 & 23.23 & 0.1 & -0.22 & 0.1 & 0.17 & -0.0 & 179.73 & 0.12  \\
2019-Aug-04 & 04:48 & 0.89 & 0.98 & 48 & 65.71 & 20.42 & 0.1 & -0.22 & 0.1 & 0.14 & 0.13 & 179.7 & 0.14  \\
2019-Aug-08 & 05:41 & 0.96 & 0.99 & 4 & 62.65 & 18.39 & 0.91 & -0.14 & 1.32 & $-$ & $-$ & 179.79 & 2.05  \\
2019-Aug-21 & 04:38 & 1.17 & 1.01 & 32 & 54.99 & 14.25 & 0.15 & -0.17 & 0.15 & 0.03 & 0.05 & 179.66 & 0.29  \\
2019-Aug-31 & 03:50 & 1.3 & 0.99 & 104 & 49.94 & 11.39 & 0.07 & -0.06 & 0.07 & -0.0 & -0.03 & 179.84 & 0.18 \\
2019-Sep-09 & 04:06 & 1.42 & 0.97 & 64 & 45.27 & 9.29 & 0.87 & -0.03 & 0.83 & 1.13 & 0.94 & 179.93 & 2.54  \\
2019-Sep-18 & 02:50 & 1.52 & 0.95 & 56 & 40.21 & 6.66 & 0.09 & 0.08 & 0.09 & -0.11 & 0.0 & 0.33 & 0.41 \\
2019-Sep-26 & 03:21 & 1.61 & 0.93 & 48 & 35.22 & 4.72 & 0.07 & -0.09 & 0.07 & 0.04 & -0.02 & 179.43 & 0.45 \\
2019-Oct-01 & 03:20 & 1.66 & 0.92 & 40 & 31.93 & 3.41 & 0.09 & -0.04 & 0.09 & 0.02 & 0.09 & 179.64 & 0.73 \\

\hline
 
\end{tabular}
\end{table*}

\begin{figure}[htb]
    \centering
    \includegraphics[width=\linewidth]{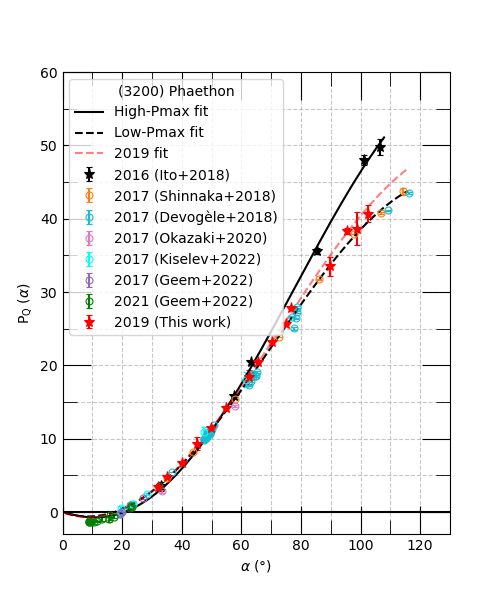}
    \caption{Polarisation of Phaethon as a function of phase angle. The low-\pmax{} polarisation phase curve is fitted using our 2019 data combined with that of \citet{shinnaka2018, devogele2018, okazaki2020, kiselev2022, geem2022}. The high-\pmax{} curve is fitted using data from \citet{ito2018}}
    \label{fig:polcurve}
\end{figure}

% ===================================================================================================

\section{Results}

\subsection{Phase angle dependence of polarisation}
\label{sec:results_phaseanglepol}
Table \ref{tablePQ} presents the nightly average of our aperture polarimetric measurements. These values are plotted as a function of phase angle in Figure \ref{fig:polcurve}, together with R-band measurements reported in previous studies. 

Previous polarimetric observations of Phaethon obtained during its 2016 and 2017 apparitions show a pronounced difference, of the order of $\sim 10$ percentage points, at large phase angles. In particular, \citet{ito2018} reported an extreme polarisation value of $P = 50\%$ at $\alpha = 106.5^{\circ}$, whereas \citet{shinnaka2018} and \citet{devogele2018} measured approximately $40\%$ at comparable phase angles. Since neither dataset show the expected decline in polarisation beyond maximum, as observed for (1566) Icarus \citep{ishiguro2017} for example, the exact value and phase angle of the polarisation maximum cannot be determined. Nevertheless, the observations clearly define distinct high- and low-\pmax{} branches.

Our 2019 measurements are evidently more consistent with the lower-polarisation branch at large phase angles during the 2017 apparition. However, most of our measurements are slightly more polarised than the 2017 observations, suggesting a possible small offset between the datasets.

We fit the high- and low-\pmax{} polarisation phase curves, as well as that of the 2019 measurements independently, using the empirical function commonly adopted for solar system bodies \citep{lumme1993, penttila2005}:
\begin{equation}
\pr(\alpha) = b \sin^{c_1}\alpha \cos^{c_2} \frac{1}{2} \alpha \sin(\alpha - \ainv)
\label{eq:trig}
\end{equation}
\noindent where $b,~c_1$, and $c_2$~are free parameters that determine the shape and amplitude of the phase curve, and $\ainv$ is the inversion angle. 

%\pr(\alpha) = b \sin^{c_1}(\alpha) \cos^{c_2} \left(\frac{1}{2} \alpha\right) \sin(\alpha - \ainv)

The low-\pmax{} branch was fitted using the published observations of \citet{shinnaka2018}, \citet{devogele2018}, \citet{okazaki2020}, \citet{kiselev2022}, and \citet{geem2022}. 
The latter study presented both new observations obtained in 2021 and re-analysis of a subset 2017 observations originally reported by \citet{shinnaka2018}. We use the re-analysed set here to avoid duplication. The high-\pmax{} branch and the independent 2019 curve were fitted using the measurements of \citet{ito2018} and this work, respectively. 

When fitted independently, the 2019 measurements yield a polarimetric phase curve that differs slightly from the combined low-\pmax{} curve. To assess whether this difference is statistically significant, we performed a Bayesian Information Criterion (BIC) analysis, following \citet{penttila2024}. Specifically, we compared models in which the 2017 and 2019 measurements are fitted either jointly with a single phase curve or independently with separate curves. Using a one-sided residual sign test, we additionally tested whether the 2019 measurements are preferentially located above the jointly fitted phase curve. The BIC analysis favours the model in which the 2017 and 2019 observations are fitted independently. Similarly, the sign test indicates that the 2019 measurements are systematically shifted toward higher polarisation compared to the jointly fitted phase curve.
Details of the statistical analysis are given in Appendix~\ref{app:stat-analysis}.

This result suggests that our 2019 observations sample a transition region between the high-\pmax{} and low-\pmax{} polarimetric phase curves of 2016 and 2017, respectively. Nevertheless, in comparison to both apparitions, the 2019 measurements are more consistent with the 2017 trend.

\begin{figure*}[htb]
    \centering
    \includegraphics[width=\linewidth]{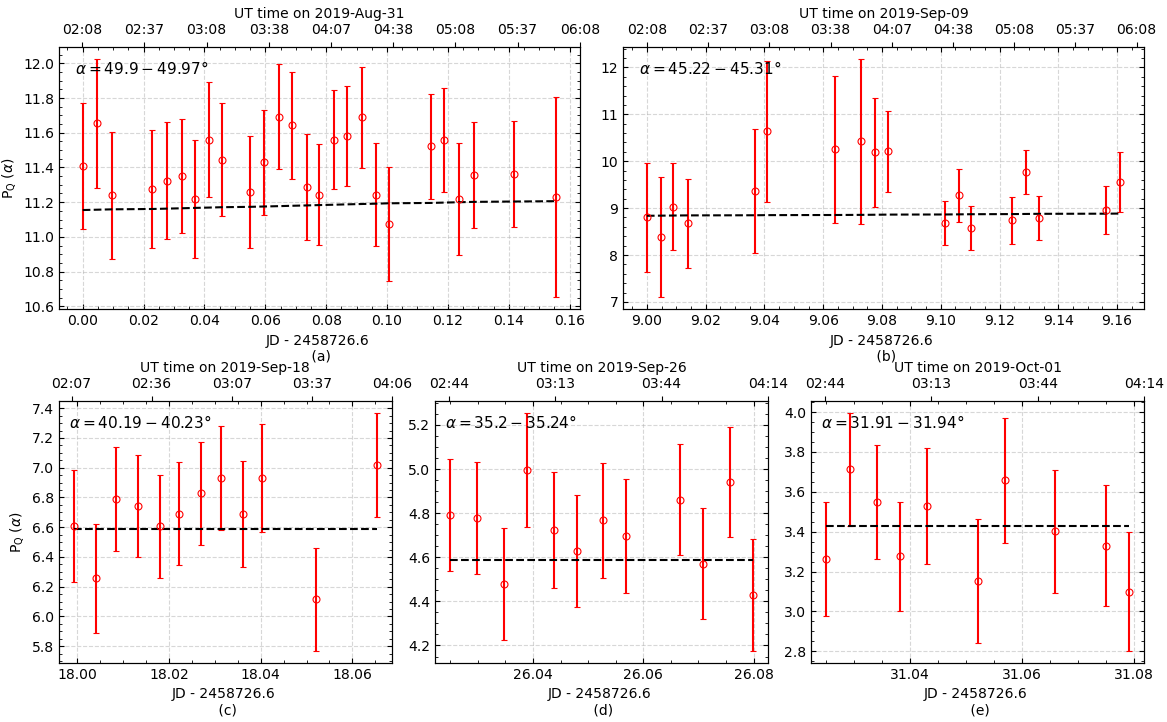}
    \caption{Time-resolved measurements of \pq\ obtained during five 2019 observing epochs. The horizontal dashed lines represent the best-fit low-\pmax{} polarimetric phase curve over the corresponding phase angle range of each epoch, indicated in each panel.}
    \label{fig:polrot}
\end{figure*}

% ===================================================================================================
\subsection{Geometric albedo}
The slope of the polarisation phase curve at the inversion angle $h$ is known to correlate empirically with the geometric albedo $p_{\text{V}}$ of asteroids \citep{geake1986}. This relation is expressed as $\log (p_{\text{V}}) = C_1 \log(h) + C_2$, where $C_1$~and $C_2$~are empirically calibrated constants. It is worth noting that these calibration constants were derived using V-band polarimetric and albedo measurements \citep{cellino2015,lupishko2018}, whereas our observations were obtained in the R-band. However, given Phaethon's approximately flat spectral slope between the V- and R-bands, \citet{geem2022} considered the V- and R-band geometric albedos to be equivalent. Therefore, we adopt $p_{\text{V}}\simeq p_{\text{R}}$.

Fitting Eq.~\ref{eq:trig} to our 2019 data gives $h=0.22\pm0.01 \%/\rm{deg}$ at $\ainv=19.97\pm0.49^{\circ}$. Published estimates of Phaethon's geometric albedo range from $\sim0.05$ to $0.16$ depending on the method used \citep[see][for a comprehensive compilation]{kiselev2022}. Using polarimetry, values of $0.05-0.06$ were derived by \citet{devogele2018} and \citet{kiselev2022}, while an intermediate value of 0.14 was derived by \citet{shinnaka2018}. Considering this range of values, we apply both calibrations of \citet{cellino2015}: the unrestricted calibration ($C_1 = -1.111, ~C_2 = -1.781$) and that calibrated using objects with $p_V \geq 0.08$ ($C_1 = -0.800, ~C_2 = -1.467$). For the calculated $h$ value, these give $p_{\text{R}}=0.098-0.12$. This value lies between the low- and high-albedo estimates reported in previous studies.

 %===================================================================================================
\subsection{Rotational dependence of polarisation}
The possibility of rotational variability in Phaethon's polarisation was first investigated by \citet{borisov2018}, who analysed a subset of the observations presented in \citet{devogele2018} and reported a variation in polarisation that they attributed to Phaethon's rotation. However, when searching for a similar signature, \citet{shinnaka2018} did not identify a periodic variation in polarisation consistent with the known rotation period in their own dataset. 

Motivated by this discrepancy, we searched for rotational variability in our 2019 observations. The two longest observation sequences, obtained on August 31 and September 9, provide approximately 4~hr of continuous polarimetric monitoring and therefore offer the best opportunity to detect variability associated with Phaethon's $\sim 3.6$~hr rotation period \citep{hanus2016}. We additionally considered the shorter time-series observations from the final three epochs.

In Fig.~\ref{fig:polrot}, we plot the time-resolved measurements of \pq\ for each epoch examined for periodic behaviour, together with the value predicted by the best-fit polarimetric phase curve over the corresponding range of phase angles. Due to the small change in phase angle during each observing sequence, the expected polarisation varies only minimally over the duration of the observations and therefore appears approximately constant. The measurements show scatter around the expected value, but no clear systematic variation with time is apparent. In particular, the two longest observing sequences, which each sample approximately one full rotation of Phaethon, do not show an obvious coherent modulation in \pq.

For a more quantitative search for periodic variability, we analysed the measurements using a Lomb--Scargle periodogram \citep{lomb1976,scargle1982}. First, we subtracted from each individual measurement the value predicted at the corresponding phase angle by the best-fitting polarimetric phase curve derived for the 2019 curve. The resulting phase angle-corrected residuals were then analysed using a Lomb--Scargle periodogram, incorporating the uncertainties of the individual measurements. We searched for periodic signals over periods between 1 and 20~hr, encompassing Phaethon's known rotational period of 3.6~hr. The periodogram does not reveal a significant peak at, or close to, the known rotational period, providing no evidence for a stable periodic modulation associated with Phaethon's rotation.

Our time-resolved measurements, therefore, do not reveal a coherent variation that persists across the observing sequences or that is consistent with Phaethon's known rotational period. The absence of a detectable rotational signal in our data is consistent with \citet{shinnaka2018}, although it does not by itself rule out rotational variability below our detection threshold. The non-detection may also reflect the specific viewing geometry sampled during the 2019 observations, as well as the limited time coverage and achieved uncertainties.

% ===================================================================================================
\subsection{Surface flux distribution}
\label{discussion:viewinggeometry}

Figure~\ref{fig:polcurve} shows that Phaethon exhibited three distinct polarimetric behaviours at large phase angles during the 2016, 2017, and 2019 apparitions, and Figs.~\ref{fig:ito}, \ref{fig:shinnaka}, and \ref{fig:not} illustrate that the viewing geometry of Phaethon varied substantially between the apparitions.

\begin{figure*}[htb]
    \centering
    \includegraphics[width=\linewidth]{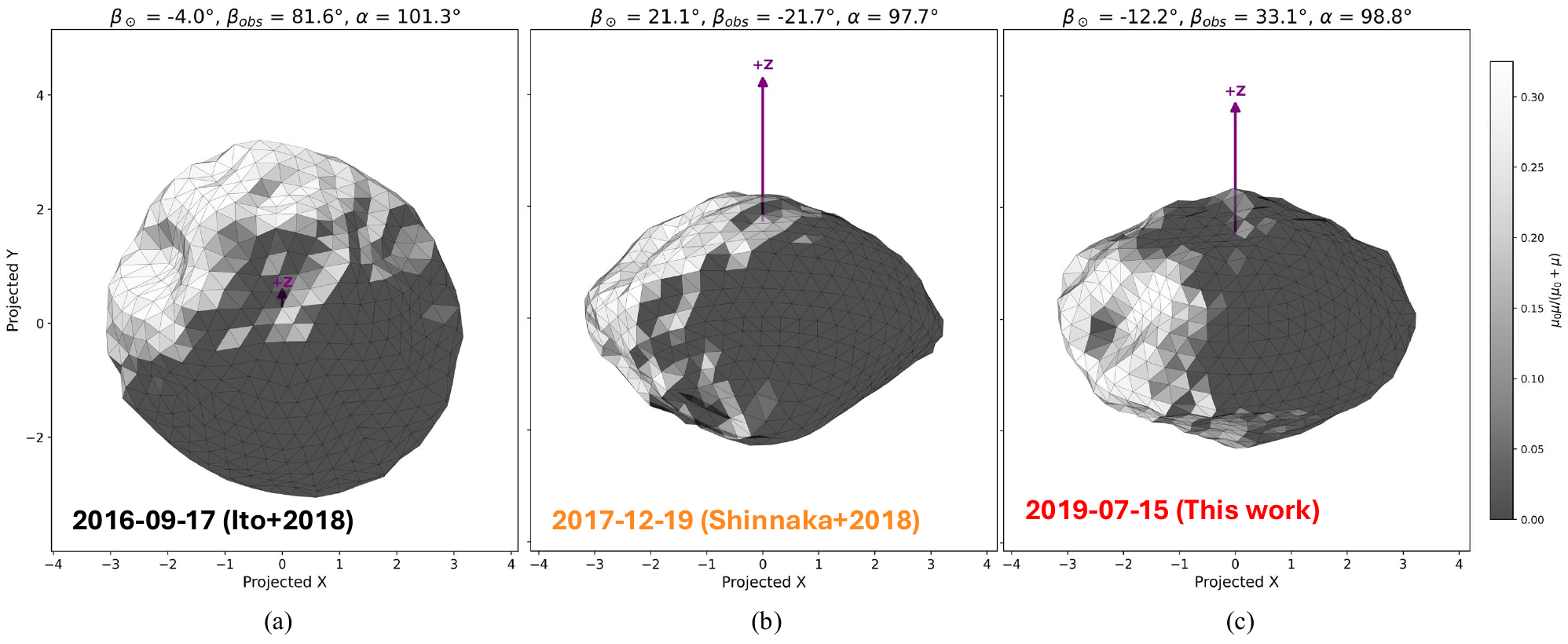}
    \caption{Rendered viewing geometry of Phaethon for epochs around $\alpha \sim 100^{\circ}$ during the (a) 2016, (b) 2017, and (c) 2019 apparitions. Facet shading represents the relative amount of reflected sunlight that is observed according to the Lommel-Seeliger model (see text for details). An arbitrary rotation phase was used and the direction of the spin pole is shown as the +z arrow.}
    \label{fig:shapemodel}
\end{figure*}

\begin{figure*}[htb]
    \centering
    \includegraphics[width=\linewidth]{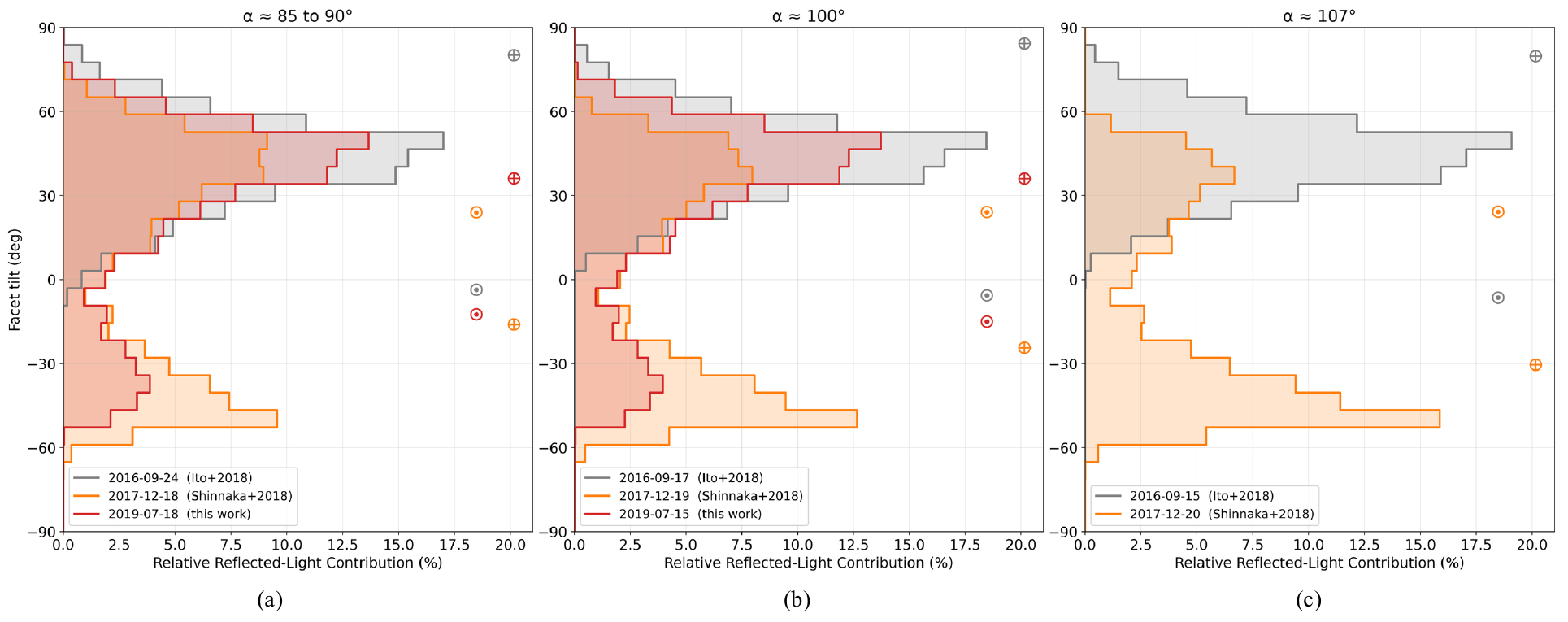}
    \caption{Reflected light distributions across Phaethon's surface for epochs around (a) $\alpha\sim85-90^{\circ}$, (b) $\alpha\sim100^{\circ}$, and (c) $\alpha\sim107^{\circ}$ during the 2016 (grey), 2017 (orange), and 2019 (red) apparitions. The sub-solar and sub-observer latitudes are denoted by the symbols $\odot$ and $\oplus$, respectively. Their numerical values, as well percentage of flux originating from positive- and negative-tilt facets are listed in Table~\ref{tab:subsunobs}.
    See text for details on how these were calculated.}
    \label{fig:reflectance}
\end{figure*}

In Figures~\ref{fig:shapemodel} and~\ref{fig:reflectance}, we directly compare the viewing geometry and flux distributions for a subset of observing epochs at phase angles $85^{\circ}\lesssim\alpha\lesssim90^{\circ}$, $\alpha\sim100^{\circ}$, and $\alpha\sim107^{\circ}$.  The corresponding sub-solar and sub-observer latitudes, together with the contributions from the positive- and negative-tilt facets, are given in Table~\ref{tab:subsunobs}. These plots show that the 2016 observations are strongly weighted towards the northern hemisphere, with the largest contribution to the observed flux originating from northern areas having facet tilts around $30^{\circ}<\beta<50^{\circ}$. Less than 1\% of flux originates from the negative-tilt facets, indicating that the observed signal is dominated almost exclusively by the northern hemisphere.

In contrast, the 2017 observations show a substantially different distribution, with contributions from both positive and negative facet tilts. At $85^\circ\lesssim\alpha\lesssim90^\circ$, the observed flux is weighted slightly towards the positive facet tilts; at $\alpha\sim100^{\circ}$, the contributions from the facet tilts are approximately balanced; while at $\alpha\sim107^{\circ}$, a larger contribution comes from the negative facet tilts. The 2019 observations are, again, dominated by the northern hemisphere, but include a significant contribution from the southern hemisphere.

Further, the three apparitions sample some surface regions that are not visible in the other epochs. In particular, the 2016 geometry samples facets with very large positive tilts ($+80^{\circ}$), correlating with the northern polar region, that are not sampled in 2017 nor 2019, while the 2017 geometry samples a region with facet tilts $-60^{\circ}$ that is not sampled in 2016 nor 2019.

Overall, the flux distributions suggest that the differences in \pmax{} between the 2016, 2017, and 2019 observations may be related to the surface regions sampled in each epoch, with higher \pmax{} generally associated with a greater contribution from the northern hemisphere.
However, the relationship is not linear.

\begin{table*}[htb]
    \centering    
    \caption{$^{\rm{a}}$Phase angle; $^{\rm{b}}$sub-solar latitude; $^{\rm{c}}$sub-observer latitude; $^{\rm{d}}$percentage of flux originating from northern hemisphere; and $^{\rm{e}}$percentage of flux originating from southern hemisphere (the remainder originates from the equator) for epochs included in Figure~\ref{fig:reflectance}.}
    \begin{tabular}{llcccc}
    \hline \hline
     $^{\rm{a}}\alpha$($^{\circ}$) & Epoch & $^{\rm{b}}\beta_{\odot} (^{\circ})$ & $^{\rm{c}}\beta_{\oplus} (^{\circ})$ & $^{\rm{d}}$N(\%) & $^{\rm{e}}$S(\%) \\
    \hline
    85--90 & 2016-09-24 (Ito+2018) & -2 & 83 & 99.01 & 0.17 \\
            & 2017-12-18 (Shinnaka+2017) & 21 & -14 & 57 & 40 \\
            & 2019-07-18 (This work) & -10 & 33 & 78.27 & 19.85  \\ \\
    100     & 2016-09-17 (Ito+2018) & -4 & 82 & 99.98 & 0.00 \\
            & 2017-12-19 (Shinnaka+2017) & 21 & -22 & 47 & 51 \\
            & 2019-07-15 (This work) & -12 & 33 & 77.72 & 20.38 \\ \\
    107     & 2016-09-15 (Ito+2018) & -5 & 77 & 99.99 & 0.00 \\
            & 2017-12-20 (Shinnaka+2017) & 21 & -27 & 38 & 60 \\
    \hline
    \end{tabular}
    \label{tab:subsunobs}
\end{table*}

% ===================================================================================================
\section{Discussion}
The 2019 observations presented here exhibit an intermediate \pmax{} behaviour between the low-\pmax{} and high-\pmax{} behaviours observed in the 2017 and 2016 apparitions, respectively, but are most similar to the 2017 observations. On the other hand, their surface-flux distributions are an intermediate case between the 2016 and 2017 apparitions, but are most similar to the 2016 case. In the following sections, we consider several factors potentially affecting the polarisation response. 

\subsection{Latitudinal heterogeneity}
% start with physics...
Let us first consider the effects of regolith grain size and albedo. For particles in the geometric optics regime, larger grain sizes typically correspond to a higher \pmax{} value \citep{frattin2019, frattin2022, munoz2021}. This can be understood in terms of the relative contributions of external reflections, which tend to increase the linear polarisation of scattered light, and multiple scattering, which tends to depolarise light. Since larger particles are more opaque, a larger fraction of light is externally reflected from the particle surfaces compared to the incident light that undergoes multiple scattering within the particles. As a consequence, for particles in the geometric optics regime, light scattered by larger particles in general results in more enhanced polarisation compared to light scattered by smaller particles \citep{Golding2023}.

% then maclennan...present logical explanation for Ito
Thermophysical modelling by \citet{maclennan2022} report heterogeneity of regolith properties across the surface, with the southern hemisphere consistent with finer particles ($<300$~{\textmu}m) and the northern hemisphere with coarser particles ($>5\,$mm). Therefore, the larger particle sizes of the northern hemisphere could provide a natural explanation for the higher-\pmax{} observed in 2016 by \citet{ito2018}, whose observations sample the northern hemisphere almost exclusively. The fact that the differences in polarisation are only found at larger phase angles also supports this hypothesis, as the contribution of large particles is most evident at large phase angles \citep{munoz2021}.

Further, it is worth noting that geometric albedo and grain size are not independent effects. The hypothesized hemispherical differences in regolith grain size distributions \citep{maclennan2022} could also produce corresponding differences in geometric albedo, in such a way that the northern hemisphere would have a lower albedo compared to the southern hemisphere. The darker northern region would, likewise, correspond to higher polarisation of scattered light, consistent with the well-known Umov effect \citep{umov1905}.

Overall, our analysis suggests that the heterogeneity of surface properties may be a plausible explanation for the higher-\pmax{} measured in 2016 by \citet{ito2018}. Quantifying these effects more precisely, however, would require detailed light-scattering modelling that simultaneously accounts for the surface properties and observing geometry.

%\subsection{Rotational heterogeneity}
%Next, we consider longitudinal differences in particle properties. Geophysical analysis suggest a depression in the northern hemisphere \citep{jo2025}, while radar observations identified a concavity near $30^\circ$ latitude that may correspond to a large crater \citep{taylor2019}. These findings suggest the presence of a localised region with anomalous surface properties. 

%If such a region had substantially difference grain size or albedo, it would be expected to produce a rotational signature as it moves into and out of view. However, rotational heterogeneity has not been consistently detected in photometric, spectroscopic and polarimetric observations \citep{kareta2018,lazzarin2019,lee2019,borisov2018,shinnaka2018,ohtsuka2020}. Our observations span  multiple rotations of Phaethon but show no variation in polarisation associated with a particular rotational phase. Thus, while a localised anomalous surface region cannot be ruled out, our observations and analysis provide no evidence that longitudinal variations in particle properties drive the observed polarimetric discrepancies. 

\begin{figure*}[htb]
    \centering
    \includegraphics[width=\linewidth]{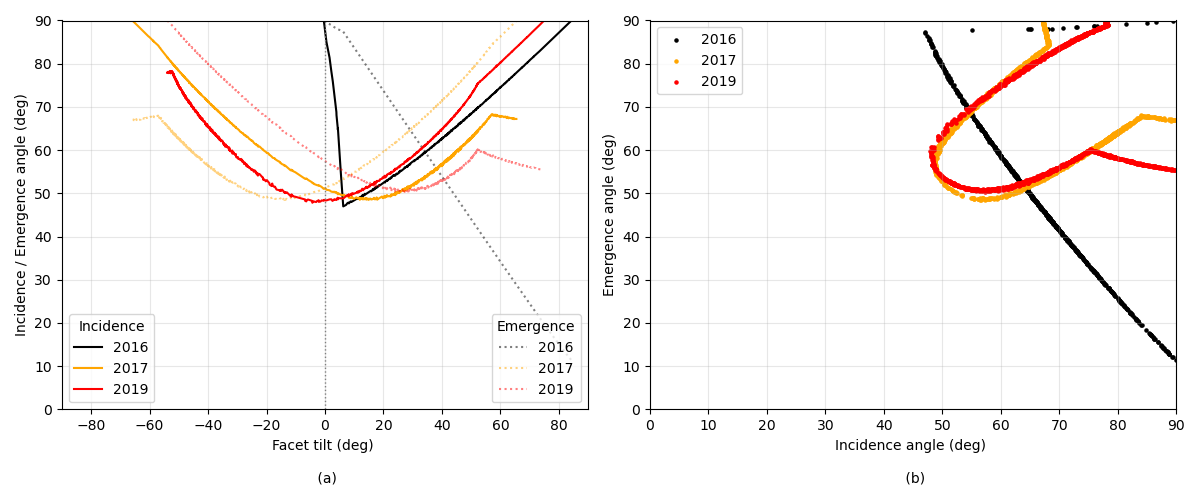}
    \caption{Rotationally averaged incidence and emergence angles across Phaethon's surface for epochs at $\alpha\sim100^{\circ}$ during the 2016, 2017, and 2019 apparitions. (a) Incidence (solid line) and emergence angles (dotted line) versus facet tilt. (b) Incidence angle versus emergence angle of all facets included in (a).}
    \label{fig:ivse}
    
\end{figure*}
\subsection{Viewing geometry}
Next, we consider viewing geometry. In addition to affecting the observability of surface facets, the sub-observer and sub-solar latitudes dictate the incidence and emergence angles of a given surface region. This, in turn, can influence its polarimetric response by altering the relative contributions of external reflections and multiple scattering. As the incidence angle increases, particles increasingly cast shadows over one another, while as the emergence angle increases, they increasingly mask one another from the observer. Thus, geometries in which both the incidence and emergence angles are large enhance the combined effects of shadowing and masking, as the surface is both illuminated and observed at more oblique angles \citep{parviainen2007}. 

Moreover, due to the effects of reciprocity, interchanging the incident and emergent angles of a given light path in a regolith produces the same scattering response, apart from the transformation of the scattering reference frame. In other words, in reciprocal light paths will results in the same scattering response, despite a large change in incidence and emergence angles. Therefore, incidence and emergence angles should not be considered independently, but together. 

In Figure~\ref{fig:ivse}, we examine the incidence and emergence angles across the rotationally averaged asteroid surface, considering facets which are simultaneously illuminated ($0^{\circ}<i<90^{\circ}$) and visible to the observer ($0^{\circ}<e<90^{\circ}$). When comparing the incident and emergence angles per facet tilt, Fig.~\ref{fig:ivse}a shows that both angles in the 2016 observations differ from those in the 2017 and 2019 observations. This distinction is also evident when comparing the incidence versus emergence angle of all facet tilts in Fig.~\ref{fig:ivse}b: the 2017 and 2019 observations follow a broadly similar relation, whereas the 2016 observations follow a markedly different relation. More specifically, in 2017 and 2019, Phaethon's surface was generally observed with more extreme incidence and emergence angles compared to that in 2016. This implies that shadowing and masking effects are more present in the 2017 and 2019 observations compared to 2016. 

The quantitative effects of shadowing and masking effects on polarisation still remains a topic of interest. Despite the 2019 observations being dominated by the northern hemisphere, where larger particles are hypothesized to be present \citet{maclennan2022}, the measured polarisation is more consistent with the lower-\pmax{} behaviour observed in 2017. This may imply that the more extreme incidence and emergence angles in 2019 compared to 2016 may act to reduce the polarisation. In other words, the enhanced polarising effect of the large particles in the northern hemisphere may be partially offset by the depolarising effects of the large incidence and emergence angles, resulting in intermediate polarisation. 

This may be understood as the following. In 2017 and 2019, the large incidence angles mean that illumination mostly reaches the tops of the surfaces of the larger regolith particles, while the large emergence and phase angles place the observer at a substantially different viewing direction from the illumination direction. The consequence of this configuration is that certain externally reflected light paths are not possible in the regolith \citep{muinonen2001,parviainen2009}. Instead, a large fraction of light that reaches the observer has undergone multiple scattering, which reduces the polarisation. Thus, this could explain why the 2017 and 2019 observations show lower polarisation than 2016 despite the contribution from the northern hemisphere.

\subsection{Perihelion related activity}  
Finally, we consider possible effects of dust ejection. A weak, short-lived $(\sim 2$~day) activity of Phaethon was discovered shortly after perihelion passage in 2009, 2012, and 2016 from near-Sun STEREO observations \citep{jewitt2010,li2013,hui2017}, and was initially attributed to the ejection of micron-sized dust \citep{jewitt2013}. Subsequent analyses of STEREO and SOHO observations instead found that the brightening is consistent with sodium fluorescence emission, without requiring scattered light from ejected dust \citep{zhang2023,hui2023}. While some material loss or regolith alteration is expected given the extreme temperatures reached at perihelion \citep{maclennan2024}, there is currently little observational evidence for substantial dust production.

Nevertheless, even a small amount of dust could potentially affect polarisation. Following the DART impact on Dimorphos, the ejecta produced a statistically significant change in the polarimetric signal \citep{gray2024PSJ,bagnulo2023,penttila2024} that persisted even after the system brightness had returned to its expected pre-impact value \citep{gray2024PSJ}. We therefore consider whether transient perihelion related dust could contribute to Phaethon's polarimetric behaviour. 

Given Phaethon's short orbital period of 1.43~yr, multiple perihelion passages occurred between the observing epochs considered here. The relevant perihelion passages, determined using JPL Horizons, occurred on 2016 August 20, 2018 January 26, and 2019 July 3. The high phase angle observations of \citet{ito2018} were obtained approximately one month after the 2016 perihelion passage, whereas \citet{shinnaka2018} and \citet{devogele2018} observed before the subsequent 2018 perihelion. Our 2019 observations began less than two weeks after the 2019 perihelion passage, placing them within a comparable post-perihelion interval to the 2016 observations. 

Despite this similar timing, the 2019 measurements do not show as high a \pmax{} as the 2016 measurements. If transient dust ejection were responsible for the enhanced polarisation observed in 2016, a similar effect might be expected in our 2019 observations. Therefore, we find no evidence that perihelion related dust ejection is responsible for the different polarimetric behaviours.  

\section{Conclusions}
We present new R-band polarimetric observations of near-Earth asteroid (3200) Phaethon obtained with the Nordic Optical Telescope during its 2019 apparition, covering the phase angle range $32^{\circ}<\alpha<103^{\circ}$. We use the latest shape model and viewing geometry during the 2016, 2017, and 2019 apparitions to investigate the surface regions contributing to the observations. With this information, we explore several possible explanations for the reported difference in polarisation. Our main results are as follows:
\begin{itemize}
    \item The 2019 observations exhibit an intermediate \pmax{} behaviour between the high- and low-\pmax{} trends observed during the 2016 and 2017 apparitions, respectively. 
    \item The best-fit polarimetric phase curve is characterised by a slope of $h=0.22 \pm 0.01\,\%/\mathrm{deg}$ at $\ainv= 19.97^{\circ} \pm 0.49^{\circ}$, corresponding to R-band geometric albedo of $p_{\text{R}} = 0.098-0.12$.
    \item We find no evidence for rotational variation in polarisation in our observations. 
    \item Modelling the surface-flux distribution show that the 2016 observations exclusively sample the northern hemisphere; 2017 observations contain significant contributions from both hemispheres; and 2019 observations are dominated by the northern hemisphere.
    \item Phaethon's surface was generally observed with more extreme incidence and emergence angles in 2017 and 2019 compared to that in 2016.
\end{itemize}

Our analysis is consistent with the interpretation of latitudinal heterogeneity across Phaethon, as proposed by \citet{maclennan2022}, whereby the larger grains in the northern hemisphere produce a higher polarisation response. At the same time, the more extreme incidence and emergence angles sampled in 2019 compared to 2016 may result in increased shadowing and masking, which could have a depolarising effect. Thus, the intermediate polarisation measured in our 2019 observations may result from the combined influence of these competing effects. 

Overall, the disk-integrated nature of the measurements and the possibly heterogeneous surface make it difficult to disentangle the effects of surface properties and viewing geometry. Detailed light-scattering modelling using realistic regolith properties will be required to determine whether these effects can reproduce the observed differences in \pmax{}. The regolith properties of Phaethon are being pursued in a complementary work (Vuori et al., in preparation). 

Future observations obtained at viewing geometries comparable to those of \citet{ito2018}, together with detailed light scattering modelling, will provide a direct test of the possible explanations considered here and will be particularly valuable for interpreting future observations of Phaethon by \textit{DESTINY$^{+}$}.

%TC:ignore
\section*{Acknowledgements}
Based on observations made with the Nordic Optical Telescope, owned in collaboration by the University of Turku and Aarhus University, and operated jointly by Aarhus University, the University of Turku and the University of Oslo, representing Denmark, Finland and Norway, the University of Iceland and Stockholm University at the Observatorio del Roque de los Muchachos, La Palma, Spain, of the Instituto de Astrofísica de Canarias. The NOT data were obtained under program ID P59-018 (PI Granvik). The data presented here were obtained in part with ALFOSC, which is provided by the Instituto de Astrofísica de Andalucía (IAA) under a joint agreement with the University of Copenhagen and NOT. 

Z.G., E.M., A.P., K.M., and M.G. acknowledge support by the Research Council of Finland (grants \#359893, \#353784 \#336546, and \#361233).
%TC:endignore

\bibliography{References}{}

@ARTICLE{Angrisani2025,
       author = {{Angrisani}, M. and {Palomba}, E. and {Longobardo}, A. and {Rognini}, E. and {Migliorini}, A. and {D'Aversa}, E. and {Pratesi}, G. and {Gisellu}, C. and {Dirri}, F. and {Piazzo}, L. and {Arai}, T.},
        title = "{Spectral Variations across Latitudes on Asteroid (3200) Phaethon}",
      journal = {\apj},
         year = 2025,
        month = feb,
       volume = {980},
       number = {1},
          eid = {127},
        pages = {127},
          doi = {10.3847/1538-4357/ada60d},
       adsurl = {https://ui.adsabs.harvard.edu/abs/2025ApJ...980..127A}
}

@ARTICLE{bagnulo2009,
   author = {{Bagnulo}, S. and {Landolfi}, M. and {Landstreet}, J.~D. and 
	{Landi Degl'Innocenti}, E. and {Fossati}, L. and {Sterzik}, M.
	},
    title = "{Stellar Spectropolarimetry with Retarder Waveplate and Beam Splitter Devices}",
  journal = {\pasp},
     year = 2009,
    month = sep,
   volume = 121,
    pages = {993-1015},
      doi = {10.1086/605654},
   adsurl = {http://adsabs.harvard.edu/abs/2009PASP..121..993B}
}

@ARTICLE{bagnulo2023,
       author = {{Bagnulo}, Stefano and {Gray}, Zuri and {Granvik}, Mikael and {Cellino}, Alberto and {Kolokolova}, Ludmilla and {Muinonen}, Karri and {Mu{\~n}oz}, Olga and {Opitom}, Cyrielle and {Penttil{\"a}}, Antti and {Snodgrass}, Colin},
        title = "{Optical Spectropolarimetry of Binary Asteroid Didymos-Dimorphos before and after the DART Impact}",
      journal = {\apjl},
         year = 2023,
        month = mar,
       volume = {945},
       number = {2},
          eid = {L38},
        pages = {L38},
          doi = {10.3847/2041-8213/acb261},
archivePrefix = {arXiv},
       eprint = {2303.11776},
 primaryClass = {astro-ph.EP},
       adsurl = {https://ui.adsabs.harvard.edu/abs/2023ApJ...945L..38B}
}

@ARTICLE{Betzler2024,
       author = {{Betzler}, A.~S. and {de Sousa}, O.~F.},
        title = "{Dual broadband phase curves of asteroid (3200) Phaethon}",
      journal = {\apss},
         year = 2024,
        month = oct,
       volume = {369},
       number = {10},
          eid = {103},
        pages = {103},
          doi = {10.1007/s10509-024-04365-9},
       adsurl = {https://ui.adsabs.harvard.edu/abs/2024Ap&SS.369..103B}
}

@ARTICLE{borisov2018,
       author = {{Borisov}, G. and {Devog{\`e}le}, M. and {Cellino}, A. and {Bagnulo}, S. and {Christou}, A. and {Bendjoya}, Ph and {Rivet}, J.-P. and {Abe}, L. and {Vernet}, D. and {Donchev}, Z. and {Krugly}, Yu and {Belskaya}, I. and {Bonev}, T. and {Steeghs}, D. and {Galloway}, D. and {Dhillon}, V. and {O'Brien}, P. and {Pollacco}, D. and {Poshyachinda}, S. and {Ramsay}, G. and {Thrane}, E. and {Ackley}, K. and {Rol}, E. and {Ulaczyk}, K. and {Cutter}, R. and {Dyer}, M.},
        title = "{Rotational variation of the linear polarization of the asteroid (3200) Phaethon as evidence for inhomogeneity in its surface properties}",
      journal = {\mnras},
         year = 2018,
        month = oct,
       volume = {480},
       number = {1},
        pages = {L131-L135},
          doi = {10.1093/mnrasl/sly140},
archivePrefix = {arXiv},
       eprint = {1807.11842},
 primaryClass = {astro-ph.EP},
       adsurl = {https://ui.adsabs.harvard.edu/abs/2018MNRAS.480L.131B}
}

@ARTICLE{cellino2015,
       author = {{Cellino}, A. and {Bagnulo}, S. and {Gil-Hutton}, R. and {Tanga}, P. and {Ca{\~n}ada-Assandri}, M. and {Tedesco}, E.~F.},
        title = "{On the calibration of the relation between geometric albedo and polarimetric properties for the asteroids}",
      journal = {\mnras},
         year = 2015,
        month = aug,
       volume = {451},
       number = {4},
        pages = {3473-3488},
          doi = {10.1093/mnras/stv1188},
archivePrefix = {arXiv},
       eprint = {1506.00554},
 primaryClass = {astro-ph.EP},
       adsurl = {https://ui.adsabs.harvard.edu/abs/2015MNRAS.451.3473C}
}

@ARTICLE{devogele2018,
       author = {{Devog{\`e}le}, M. and {Cellino}, A. and {Borisov}, G. and {Bendjoya}, Ph and {Rivet}, J. -P. and {Abe}, L. and {Bagnulo}, S. and {Christou}, A. and {Vernet}, D. and {Donchev}, Z. and {Belskaya}, I. and {Bonev}, T. and {Krugly}, Yu N.},
        title = "{The phase-polarization curve of asteroid (3200) Phaethon$^{{\textdagger}}$}",
      journal = {\mnras},
         year = 2018,
        month = sep,
       volume = {479},
       number = {3},
        pages = {3498-3508},
          doi = {10.1093/mnras/sty1587},
archivePrefix = {arXiv},
       eprint = {1806.05277},
 primaryClass = {astro-ph.EP},
       adsurl = {https://ui.adsabs.harvard.edu/abs/2018MNRAS.479.3498D}
}

@ARTICLE{frattin2019,
       author = {{Frattin}, E. and {Mu{\~n}oz}, O. and {Moreno}, F. and {Nava}, J. and {Escobar-Cerezo}, J. and {Gomez Martin}, J.~C. and {Guirado}, D. and {Cellino}, A. and {Coll}, P. and {Raulin}, F. and {Bertini}, I. and {Cremonese}, G. and {Lazzarin}, M. and {Naletto}, G. and {La Forgia}, F.},
        title = "{Experimental phase function and degree of linear polarization of cometary dust analogues}",
      journal = {\mnras},
         year = 2019,
        month = apr,
       volume = {484},
       number = {2},
        pages = {2198-2211},
          doi = {10.1093/mnras/stz129},
archivePrefix = {arXiv},
       eprint = {1901.05975},
 primaryClass = {astro-ph.EP},
       adsurl = {https://ui.adsabs.harvard.edu/abs/2019MNRAS.484.2198F}
}

@ARTICLE{frattin2022,
       author = {{Frattin}, E. and {Martikainen}, J. and {Mu{\~n}oz}, O. and {G{\'o}mez-Mart{\'\i}n}, J.~C. and {Jardiel}, T. and {Cellino}, A. and {Libourel}, G. and {Muinonen}, K. and {Peiteado}, M. and {Tanga}, P.},
        title = "{Experimental phase function and degree of linear polarization curve of olivine and spinel and the origin of the Barbarian polarization behaviour}",
      journal = {\mnras},
         year = 2022,
        month = dec,
       volume = {517},
       number = {4},
        pages = {5463-5472},
          doi = {10.1093/mnras/stac2895},
archivePrefix = {arXiv},
       eprint = {2306.11358},
 primaryClass = {astro-ph.EP},
       adsurl = {https://ui.adsabs.harvard.edu/abs/2022MNRAS.517.5463F}
}

@ARTICLE{geake1986,
       author = {{Geake}, J.~E. and {Dollfus}, A.},
        title = "{Planetary surface texture and albedo from parameter plots of optical polarization data}",
      journal = {\mnras},
         year = 1986,
        month = jan,
       volume = {218},
        pages = {75-91},
          doi = {10.1093/mnras/218.1.75},
       adsurl = {https://ui.adsabs.harvard.edu/abs/1986MNRAS.218...75G}
}

@ARTICLE{geem2022,
       author = {{Geem}, Jooyeon and {Ishiguro}, Masateru and {Takahashi}, Jun and {Akitaya}, Hiroshi and {Kawabata}, Koji S. and {Nakaoka}, Tatsuya and {Imazawa}, Ryo and {Mori}, Fumiki and {Jin}, Sunho and {Bach}, Yoonsoo P. and {Jo}, Hangbin and {Kuroda}, Daisuke and {Hasegawa}, Sunao and {Yoshida}, Fumi and {Ishibashi}, Ko and {Sekiguchi}, Tomohiko and {Beniyama}, Jin and {Arai}, Tomoko and {Ikeda}, Yuji and {Shinnaka}, Yoshiharu and {Granvik}, Mikael and {Siltala}, Lauri and {Djupvik}, Anlaug A. and {Kasikov}, Anni and {Pinter}, Viktoria and {Knudstrup}, Emil},
        title = "{(3200) Phaethon polarimetry in the negative branch: new evidence for the anhydrous nature of the DESTINY$^{+}$ target asteroid}",
      journal = {\mnras},
         year = 2022,
        month = oct,
       volume = {516},
       number = {1},
        pages = {L53-L57},
          doi = {10.1093/mnrasl/slac072},
archivePrefix = {arXiv},
       eprint = {2208.11912},
 primaryClass = {astro-ph.EP},
       adsurl = {https://ui.adsabs.harvard.edu/abs/2022MNRAS.516L..53G}
}

@ARTICLE{Golding2023,
       author = {{Golding}, Rachel M. and {Lapszynski}, Christopher S. and {Bachmann}, Charles M. and {Lee}, Chris H.},
        title = "{The Effect of Grain Size on Hyperspectral Polarization Data of Particulate Material}",
      journal = {Remote Sensing},
         year = 2023,
        month = jul,
       volume = {15},
       number = {14},
          eid = {3668},
        pages = {3668},
          doi = {10.3390/rs15143668},
       adsurl = {https://ui.adsabs.harvard.edu/abs/2023RemS...15.3668G}
}

@ARTICLE{granvik2016,
       author = {{Granvik}, Mikael and {Morbidelli}, Alessandro and {Jedicke}, Robert and {Bolin}, Bryce and {Bottke}, William F. and {Beshore}, Edward and {Vokrouhlick{\'y}}, David and {Delb{\`o}}, Marco and {Michel}, Patrick},
        title = "{Super-catastrophic disruption of asteroids at small perihelion distances}",
      journal = {\nat},
         year = 2016,
        month = feb,
       volume = {530},
       number = {7590},
        pages = {303-306},
          doi = {10.1038/nature16934},
       adsurl = {https://ui.adsabs.harvard.edu/abs/2016Natur.530..303G}
}

@ARTICLE{gray2024PSJ,
       author = {{Gray}, Zuri and {Bagnulo}, Stefano and {Granvik}, Mikael and {Cellino}, Alberto and {Jones}, Geraint H. and {Kolokolova}, Ludmilla and {Moreno}, Fernando and {Muinonen}, Karri and {Mu{\~n}oz}, Olga and {Opitom}, Cyrielle and {Penttil{\"a}}, Antti and {Snodgrass}, Colin},
        title = "{Polarimetry of Didymos{\textendash}Dimorphos: Unexpected Long-term Effects of the DART Impact}",
      journal = {\psj},
         year = {2024},
        month = jan,
       volume = {5},
       number = {1},
          eid = {18},
        pages = {18},
          doi = {10.3847/PSJ/ad0f18},
archivePrefix = {arXiv},
       eprint = {2311.13483},
 primaryClass = {astro-ph.EP},
       adsurl = {https://ui.adsabs.harvard.edu/abs/2024PSJ.....5...18G}
}

@ARTICLE{hanus2016,
       author = {{Hanu{\v{s}}}, J. and {Delbo'}, M. and {Vokrouhlick{\'y}}, D. and {Pravec}, P. and {Emery}, J.~P. and {Al{\'\i}-Lagoa}, V. and {Bolin}, B. and {Devog{\`e}le}, M. and {Dyvig}, R. and {Gal{\'a}d}, A. and {Jedicke}, R. and {Korno{\v{s}}}, L. and {Ku{\v{s}}nir{\'a}k}, P. and {Licandro}, J. and {Reddy}, V. and {Rivet}, J.-P. and {Vil{\'a}gi}, J. and {Warner}, B.~D.},
        title = "{Near-Earth asteroid (3200) Phaethon: Characterization of its orbit, spin state, and thermophysical parameters}",
      journal = {\aap},
         year = 2016,
        month = jul,
       volume = {592},
          eid = {A34},
        pages = {A34},
          doi = {10.1051/0004-6361/201628666},
archivePrefix = {arXiv},
       eprint = {1605.05205},
 primaryClass = {astro-ph.EP},
       adsurl = {https://ui.adsabs.harvard.edu/abs/2016A&A...592A..34H}
}

@ARTICLE{hui2017,
       author = {{Hui}, Man-To and {Li}, Jing},
        title = "{Resurrection of (3200) Phaethon in 2016}",
      journal = {\aj},
         year = 2017,
        month = jan,
       volume = {153},
       number = {1},
          eid = {23},
        pages = {23},
          doi = {10.3847/1538-3881/153/1/23},
archivePrefix = {arXiv},
       eprint = {1611.07061},
 primaryClass = {astro-ph.EP},
       adsurl = {https://ui.adsabs.harvard.edu/abs/2017AJ....153...23H}
}

@ARTICLE{hui2023,
       author = {{Hui}, Man-To},
        title = "{Perihelion Activity of (3200) Phaethon is Not Dusty: Evidence from STEREO/COR2 Observations}",
      journal = {\aj},
         year = 2023,
        month = mar,
       volume = {165},
       number = {3},
          eid = {94},
        pages = {94},
          doi = {10.3847/1538-3881/acae9c},
archivePrefix = {arXiv},
       eprint = {2206.09704},
 primaryClass = {astro-ph.EP},
       adsurl = {https://ui.adsabs.harvard.edu/abs/2023AJ....165...94H}
}

@ARTICLE{ishiguro2017,
       author = {{Ishiguro}, Masateru and {Kuroda}, Daisuke and {Watanabe}, Makoto and {Bach}, Yoonsoo P. and {Kim}, Jooyeon and {Lee}, Mingyeong and {Sekiguchi}, Tomohiko and {Naito}, Hiroyuki and {Ohtsuka}, Katsuhito and {Hanayama}, Hidekazu and {Hasegawa}, Sunao and {Usui}, Fumihiko and {Urakawa}, Seitaro and {Imai}, Masataka and {Sato}, Mitsuteru and {Kuramoto}, Kiyoshi},
        title = "{Polarimetric Study of Near-Earth Asteroid (1566) Icarus}",
      journal = {\aj},
         year = 2017,
        month = nov,
       volume = {154},
       number = {5},
          eid = {180},
        pages = {180},
          doi = {10.3847/1538-3881/aa8b1a},
archivePrefix = {arXiv},
       eprint = {1709.01603},
 primaryClass = {astro-ph.EP},
       adsurl = {https://ui.adsabs.harvard.edu/abs/2017AJ....154..180I}
}

@ARTICLE{ito2018,
       author = {{Ito}, Takashi and {Ishiguro}, Masateru and {Arai}, Tomoko and {Imai}, Masataka and {Sekiguchi}, Tomohiko and {Bach}, Yoonsoo P. and {Kwon}, Yuna G. and {Kobayashi}, Masanori and {Ishimaru}, Ryo and {Naito}, Hiroyuki and {Watanabe}, Makoto and {Kuramoto}, Kiyoshi},
        title = "{Extremely strong polarization of an active asteroid (3200) Phaethon}",
      journal = {Nature Communications},
         year = 2018,
        month = jun,
       volume = {9},
          eid = {2486},
        pages = {2486},
          doi = {10.1038/s41467-018-04727-2},
       adsurl = {https://ui.adsabs.harvard.edu/abs/2018NatCo...9.2486I}
}

@ARTICLE{jewitt2006,
       author = {{Jewitt}, David and {Hsieh}, Henry},
        title = "{Physical Observations of 2005 UD: A Mini-Phaethon}",
      journal = {\aj},
         year = 2006,
        month = oct,
       volume = {132},
       number = {4},
        pages = {1624-1629},
          doi = {10.1086/507483},
       adsurl = {https://ui.adsabs.harvard.edu/abs/2006AJ....132.1624J}
}

@ARTICLE{jewitt2010,
       author = {{Jewitt}, David and {Li}, Jing},
        title = "{Activity in Geminid Parent (3200) Phaethon}",
      journal = {\aj},
         year = 2010,
        month = nov,
       volume = {140},
       number = {5},
        pages = {1519-1527},
          doi = {10.1088/0004-6256/140/5/1519},
archivePrefix = {arXiv},
       eprint = {1009.2710},
 primaryClass = {astro-ph.EP},
       adsurl = {https://ui.adsabs.harvard.edu/abs/2010AJ....140.1519J}
}

@ARTICLE{jewitt2013,
       author = {{Jewitt}, David and {Li}, Jing and {Agarwal}, Jessica},
        title = "{The Dust Tail of Asteroid (3200) Phaethon}",
      journal = {\apjl},
         year = 2013,
        month = jul,
       volume = {771},
       number = {2},
          eid = {L36},
        pages = {L36},
          doi = {10.1088/2041-8205/771/2/L36},
archivePrefix = {arXiv},
       eprint = {1306.3741},
 primaryClass = {astro-ph.EP},
       adsurl = {https://ui.adsabs.harvard.edu/abs/2013ApJ...771L..36J}
}

@ARTICLE{jo2025,
       author = {{Jo}, Hangbin and {Ishiguro}, Masateru and {Richardson}, Derek C. and {Marshall}, Sean E. and {Arai}, Tomoko and {Ishibashi}, Ko},
        title = "{The surface dynamics and geophysical environment of asteroid (3200) Phaethon}",
      journal = {\aap},
         year = 2025,
        month = apr,
       volume = {696},
          eid = {A34},
        pages = {A34},
          doi = {10.1051/0004-6361/202453408},
archivePrefix = {arXiv},
       eprint = {2502.09916},
 primaryClass = {astro-ph.EP},
       adsurl = {https://ui.adsabs.harvard.edu/abs/2025A&A...696A..34J}
}

@ARTICLE{kareta2018,
       author = {{Kareta}, Theodore and {Reddy}, Vishnu and {Hergenrother}, Carl and {Lauretta}, Dante S. and {Arai}, Tomoko and {Takir}, Driss and {Sanchez}, Juan and {Hanu{\v{s}}}, Josef},
        title = "{Rotationally Resolved Spectroscopic Characterization of Near-Earth Object (3200) Phaethon}",
      journal = {\aj},
         year = 2018,
        month = dec,
       volume = {156},
       number = {6},
          eid = {287},
        pages = {287},
          doi = {10.3847/1538-3881/aaeb8a},
archivePrefix = {arXiv},
       eprint = {1810.11157},
 primaryClass = {astro-ph.EP},
       adsurl = {https://ui.adsabs.harvard.edu/abs/2018AJ....156..287K}
}

@ARTICLE{kinoshita2007,
       author = {{Kinoshita}, D. and {Ohtsuka}, K. and {Sekiguchi}, T. and {Watanabe}, J. and {Ito}, T. and {Arakida}, H. and {Kasuga}, T. and {Miyasaka}, S. and {Nakamura}, R. and {Lin}, H.-C.},
        title = "{Surface heterogeneity of 2005 UD from photometric observations}",
      journal = {\aap},
         year = 2007,
        month = may,
       volume = {466},
       number = {3},
        pages = {1153-1158},
          doi = {10.1051/0004-6361:20066276},
       adsurl = {https://ui.adsabs.harvard.edu/abs/2007A&A...466.1153K}
}

@ARTICLE{kiselev2022,
       author = {{Kiselev}, N.~N. and {Rosenbush}, V.~K. and {Petrov}, D. and {Luk'yanyk}, I.~V. and {Ivanova}, O.~V. and {Pit}, N.~V. and {Antoniuk}, K.~A. and {Afanasiev}, V.~L.},
        title = "{Asteroid (3200) Phaethon: results of polarimetric, photometric, and spectral observations}",
      journal = {\mnras},
         year = 2022,
        month = aug,
       volume = {514},
       number = {4},
        pages = {4861-4875},
          doi = {10.1093/mnras/stac1559},
archivePrefix = {arXiv},
       eprint = {2206.00911},
 primaryClass = {astro-ph.EP},
       adsurl = {https://ui.adsabs.harvard.edu/abs/2022MNRAS.514.4861K}
}

@ARTICLE{lazzarin2019,
       author = {{Lazzarin}, M. and {Petropoulou}, V. and {Bertini}, I. and {La Forgia}, F. and {Ochner}, P. and {Migliorini}, A. and {Siviero}, A.},
        title = "{Phaethon variability during December 2017 closest approach to Earth}",
      journal = {\planss},
         year = 2019,
        month = jan,
       volume = {165},
        pages = {115-123},
          doi = {10.1016/j.pss.2018.11.006},
       adsurl = {https://ui.adsabs.harvard.edu/abs/2019P&SS..165..115L}
}

@ARTICLE{lee2019,
       author = {{Lee}, H.-J. and {Kim}, M.-J. and {Kim}, D.-H. and {Moon}, H.-K. and {Choi}, Y.-J. and {Kim}, C.-H. and {Lee}, B.-C. and {Yoshida}, F. and {Roh}, D.-G. and {Seo}, H.},
        title = "{Investigation of surface homogeneity of (3200) Phaethon}",
      journal = {\planss},
         year = 2019,
        month = jan,
       volume = {165},
        pages = {296-302},
          doi = {10.1016/j.pss.2018.12.001},
archivePrefix = {arXiv},
       eprint = {1812.01851},
 primaryClass = {astro-ph.EP},
       adsurl = {https://ui.adsabs.harvard.edu/abs/2019P&SS..165..296L}
}

@ARTICLE{li2013,
       author = {{Li}, Jing and {Jewitt}, David},
        title = "{Recurrent Perihelion Activity in (3200) Phaethon}",
      journal = {\aj},
         year = 2013,
        month = jun,
       volume = {145},
       number = {6},
          eid = {154},
        pages = {154},
          doi = {10.1088/0004-6256/145/6/154},
archivePrefix = {arXiv},
       eprint = {1304.1430},
 primaryClass = {astro-ph.EP},
       adsurl = {https://ui.adsabs.harvard.edu/abs/2013AJ....145..154L}
}

@ARTICLE{lomb1976,
       author = {{Lomb}, N.~R.},
        title = "{Least-Squares Frequency Analysis of Unequally Spaced Data}",
      journal = {\apss},
         year = 1976,
        month = feb,
       volume = {39},
       number = {2},
        pages = {447-462},
          doi = {10.1007/BF00648343},
       adsurl = {https://ui.adsabs.harvard.edu/abs/1976Ap&SS..39..447L}
}

@ARTICLE{lumme1981,
       author = {{Lumme}, K. and {Bowell}, E.},
        title = "{Radiative transfer in the surfaces of atmosphereless bodies. I. Theory.}",
      journal = {\aj},
         year = 1981,
        month = nov,
       volume = {86},
        pages = {1694-1721},
          doi = {10.1086/113054},
       adsurl = {https://ui.adsabs.harvard.edu/abs/1981AJ.....86.1694L}
}

@INPROCEEDINGS{lumme1993,
       author = {{Lumme}, K. and {Muinonen}, K.~O.},
        title = "{A Two-Parameter System for Linear Polarization of Some Solar System Objects}",
    booktitle = {Asteroids, Comets, Meteors 1993},
         year = 1993,
       editor = {{LPI Editorial Board}},
       series = {LPI Contributions},
       volume = {810},
        month = jan,
        pages = {194},
       adsurl = {https://ui.adsabs.harvard.edu/abs/1993LPICo.810..194L}
}

@ARTICLE{lupishko2018,
       author = {{Lupishko}, D.~F.},
        title = "{Generalized Calibration of the Polarimetric Albedo Scale of Asteroids}",
      journal = {Solar System Research},
         year = 2018,
        month = mar,
       volume = {52},
       number = {2},
        pages = {98-114},
          doi = {10.1134/S0038094618010069},
       adsurl = {https://ui.adsabs.harvard.edu/abs/2018SoSyR..52...98L}
}

@ARTICLE{maclennan2022,
       author = {{MacLennan}, Eric and {Marshall}, Sean and {Granvik}, Mikael},
        title = "{Evidence of surface heterogeneity on active asteroid (3200) Phaethon}",
      journal = {\icarus},
         year = 2022,
        month = dec,
       volume = {388},
          eid = {115226},
        pages = {115226},
          doi = {10.1016/j.icarus.2022.115226},
archivePrefix = {arXiv},
       eprint = {2203.08865},
 primaryClass = {astro-ph.EP},
       adsurl = {https://ui.adsabs.harvard.edu/abs/2022Icar..38815226M}
}

@ARTICLE{maclennan2024,
       author = {{MacLennan}, Eric and {Granvik}, Mikael},
        title = "{Thermal decomposition as the activity driver of near-Earth asteroid (3200) Phaethon}",
      journal = {Nature Astronomy},
         year = 2024,
        month = jan,
       volume = {8},
       number = {1},
        pages = {60-68},
          doi = {10.1038/s41550-023-02091-w},
archivePrefix = {arXiv},
       eprint = {2207.08968},
 primaryClass = {astro-ph.EP},
       adsurl = {https://ui.adsabs.harvard.edu/abs/2024NatAs...8...60M}
}

@INPROCEEDINGS{marshall_acm2026,
       author = {{Marshall}, Sean and {Ferrais}, Marin and {Rommel}, Flavia and {Inosencio}, Xavier and {McGraw}, Allison},
       title = "{Shape Models of Near-Earth Asteroids 1036 Ganymed, 3200 Phaethon, 5189 (1990 UQ), 285331 (1999 FN53), and 385186 (1994 AW1)}",
    booktitle = {Asteroids, Comets, Meteors Conference},
         year = 2026,
          url = {acm2026.eu},
       Xseries = {LPI Contributions},
       Xvolume = {2851},
        month = jul,
          eid = {200},
        Xpages = {2435}
        }

@ARTICLE{muinonen2001,
       author = {{Muinonen}, K. and {Stankevich}, D. and {Shkuratov}, Yu. G. and {Kaasalainen}, S. and {Piironen}, J.},
        title = "{Shadowing effect in clusters of opaque spherical particles}",
      journal = {\jqsrt},
         year = 2001,
        month = aug,
       volume = {70},
        pages = {787-810},
          doi = {10.1016/S0022-4073(01)00046-2},
       adsurl = {https://ui.adsabs.harvard.edu/abs/2001JQSRT..70..787M}
}

@ARTICLE{munoz2021,
       author = {{Mu{\~n}oz}, O. and {Frattin}, E. and {Jardiel}, T. and {G{\'o}mez-Mart{\'\i}n}, J.~C. and {Moreno}, F. and {Ramos}, J.~L. and {Guirado}, D. and {Peiteado}, M. and {Caballero}, A.~C. and {Milli}, J. and {M{\'e}nard}, F.},
        title = "{Retrieving Dust Grain Sizes from Photopolarimetry: An Experimental Approach}",
      journal = {\apjs},
         year = 2021,
        month = sep,
       volume = {256},
       number = {1},
          eid = {17},
        pages = {17},
          doi = {10.3847/1538-4365/ac0efa},
archivePrefix = {arXiv},
       eprint = {2109.05764},
 primaryClass = {astro-ph.EP},
       adsurl = {https://ui.adsabs.harvard.edu/abs/2021ApJS..256...17M}
}

@ARTICLE{ohtsuka2020,
       author = {{Ohtsuka}, Katsuhito and {Ito}, Takashi and {Kinoshita}, Daisuke and {Abe}, Shinsuke and {Sawai}, Kyosuke and {Funabashi}, Kazuhiro and {Kato}, Ryo and {Miyasaka}, Seidai and {Hasegawa}, Sunao and {Nakamura}, Tomoki and {Chen}, Wen-Ping},
        title = "{Full rotationally phase-resolved visible reflectance spectroscopy of 3200 Phaethon}",
      journal = {\planss},
         year = 2020,
        month = oct,
       volume = {191},
          eid = {104940},
        pages = {104940},
          doi = {10.1016/j.pss.2020.104940},
archivePrefix = {arXiv},
       eprint = {1703.00296},
 primaryClass = {astro-ph.EP},
       adsurl = {https://ui.adsabs.harvard.edu/abs/2020P&SS..19104940O}
}

@ARTICLE{ohtsuka2006,
       author = {{Ohtsuka}, K. and {Sekiguchi}, T. and {Kinoshita}, D. and {Watanabe}, J.-I. and {Ito}, T. and {Arakida}, H. and {Kasuga}, T.},
        title = "{Apollo asteroid 2005 UD: split nucleus of (3200) Phaethon?}",
      journal = {\aap},
         year = 2006,
        month = may,
       volume = {450},
       number = {3},
        pages = {L25-L28},
          doi = {10.1051/0004-6361:200600022},
       adsurl = {https://ui.adsabs.harvard.edu/abs/2006A&A...450L..25O}
}

@ARTICLE{okazaki2020,
       author = {{Okazaki}, Ryo and {Sekiguchi}, Tomohiko and {Ishiguro}, Masateru and {Naito}, Hiroyuki and {Urakawa}, Seitaro and {Imai}, Masataka and {Ono}, Tatsuharu and {Warner}, Brian D. and {Watanabe}, Makoto},
        title = "{Polarimetric and photometric observations of NEAs; (422699) 2000 PD3 and (3200) Phaethon with the 1.6m Pirka telescope}",
      journal = {\planss},
         year = 2020,
        month = jan,
       volume = {180},
          eid = {104774},
        pages = {104774},
          doi = {10.1016/j.pss.2019.104774},
archivePrefix = {arXiv},
       eprint = {1910.04305},
 primaryClass = {astro-ph.EP},
       adsurl = {https://ui.adsabs.harvard.edu/abs/2020P&SS..18004774O}
}

@ARTICLE{parviainen2007,
       author = {{Parviainen}, Hannu and {Muinonen}, Karri},
        title = "{Rough-surface shadowing of self-affine random rough surfaces}",
      journal = {\jqsrt},
         year = 2007,
        month = jul,
       volume = {106},
       number = {1-3},
        pages = {398-416},
          doi = {10.1016/j.jqsrt.2007.01.025},
archivePrefix = {arXiv},
       eprint = {2010.03927},
 primaryClass = {physics.optics},
       adsurl = {https://ui.adsabs.harvard.edu/abs/2007JQSRT.106..398P}
}

@ARTICLE{parviainen2009,
       author = {{Parviainen}, Hannu and {Muinonen}, Karri},
        title = "{Bidirectional reflectance of rough particulate media: Ray-tracing solution}",
      journal = {\jqsrt},
         year = 2009,
        month = sep,
       volume = {110},
        pages = {1418-1440},
       adsurl = {https://ui.adsabs.harvard.edu/abs/2009JQSRT.110.1418P}
}

@ARTICLE{penttila2005,
       author = {{Penttil{\"a}}, A. and {Lumme}, K. and {Hadamcic}, E. and {Levasseur-Regourd}, A.-C.},
       title = "{Statistical analysis of asteroidal and cometary polarization phase curves }",
      journal = {\aa}, 
         year = 2005,
       volume = {432},
        pages = {1081-1090},
          doi = {10.1051/0004-6361:20042133}
}

@ARTICLE{penttila2024,
       author = {{Penttil{\"a}}, Antti and {Muinonen}, Karri and {Granvik}, Mikael and {Gray}, Zuri and {Bagnulo}, Stefano and {Kolokolova}, Ludmilla and {Moreno}, Fernando},
        title = "{Modeling Linear Polarization of the Didymos─Dimorphos System before and after the DART Impact}",
      journal = {\psj},
         year = 2024,
        month = jan,
       volume = {5},
       number = {1},
          eid = {27},
        pages = {27},
          doi = {10.3847/PSJ/ad1757},
       adsurl = {https://ui.adsabs.harvard.edu/abs/2024PSJ.....5...27P}
}

@ARTICLE{scargle1982,
       author = {{Scargle}, J.~D.},
        title = "{Studies in astronomical time series analysis. II. Statistical aspects of spectral analysis of unevenly spaced data.}",
      journal = {\apj},
         year = 1982,
        month = dec,
       volume = {263},
        pages = {835-853},
          doi = {10.1086/160554},
       adsurl = {https://ui.adsabs.harvard.edu/abs/1982ApJ...263..835S}
}

@ARTICLE{shinnaka2018,
       author = {{Shinnaka}, Yoshiharu and {Kasuga}, Toshihiro and {Furusho}, Reiko and {Boice}, Daniel C. and {Terai}, Tsuyoshi and {Noda}, Hirotomo and {Namiki}, Noriyuki and {Watanabe}, Jun-ichi},
        title = "{Inversion Angle of Phase-polarization Curve of Near-Earth Asteroid (3200) Phaethon}",
      journal = {\apjl},
         year = 2018,
        month = sep,
       volume = {864},
       number = {2},
          eid = {L33},
        pages = {L33},
          doi = {10.3847/2041-8213/aadb3d},
archivePrefix = {arXiv},
       eprint = {1808.05746},
 primaryClass = {astro-ph.EP},
       adsurl = {https://ui.adsabs.harvard.edu/abs/2018ApJ...864L..33S}
}

@ARTICLE{taylor2019,
       author = {{Taylor}, Patrick A. and {Rivera-Valent{\'\i}n}, Edgard G. and {Benner}, Lance A.~M. and {Marshall}, Sean E. and {Virkki}, Anne K. and {Venditti}, Flaviane C.~F. and {Zambrano-Marin}, Luisa F. and {Bhiravarasu}, Sriram S. and {Aponte-Hernandez}, Betzaida and {Rodriguez Sanchez-Vahamonde}, Carolina and {Giorgini}, Jon D.},
        title = "{Arecibo radar observations of near-Earth asteroid (3200) Phaethon during the 2017 apparition}",
      journal = {\planss},
         year = 2019,
        month = mar,
       volume = {167},
        pages = {1-8},
          doi = {10.1016/j.pss.2019.01.009},
archivePrefix = {arXiv},
       eprint = {1902.00969},
 primaryClass = {astro-ph.EP},
       adsurl = {https://ui.adsabs.harvard.edu/abs/2019P&SS..167....1T}
}

@article{umov1905,
  author  = {Umov, N. A.},
  title   = {Chromatische Depolarisation durch Lichtzerstreuung},
  journal = {Physikalische Zeitschrift},
  volume  = {6},
  pages   = {674--676},
  year    = {1905}
}

@ARTICLE{zhang2023,
       author = {{Zhang}, Qicheng and {Battams}, Karl and {Ye}, Quanzhi and {Knight}, Matthew M. and {Schmidt}, Carl A.},
        title = "{Sodium Brightening of (3200) Phaethon near Perihelion}",
      journal = {\psj},
         year = 2023,
        month = apr,
       volume = {4},
       number = {4},
          eid = {70},
        pages = {70},
          doi = {10.3847/PSJ/acc866},
archivePrefix = {arXiv},
       eprint = {2303.17625},
 primaryClass = {astro-ph.EP},
       adsurl = {https://ui.adsabs.harvard.edu/abs/2023PSJ.....4...70Z}
}
\bibliographystyle{aasjournal}

\appendix

\setcounter{figure}{0}
\renewcommand{\thefigure}{A\arabic{figure}}

\setcounter{table}{0}
\renewcommand{\thetable}{B\arabic{table}}

\section{Reflected Sunlight Distributions}
\label{app:flux-distributions}

\begin{figure}[!ht]
    \centering
    \includegraphics[width=1\linewidth]{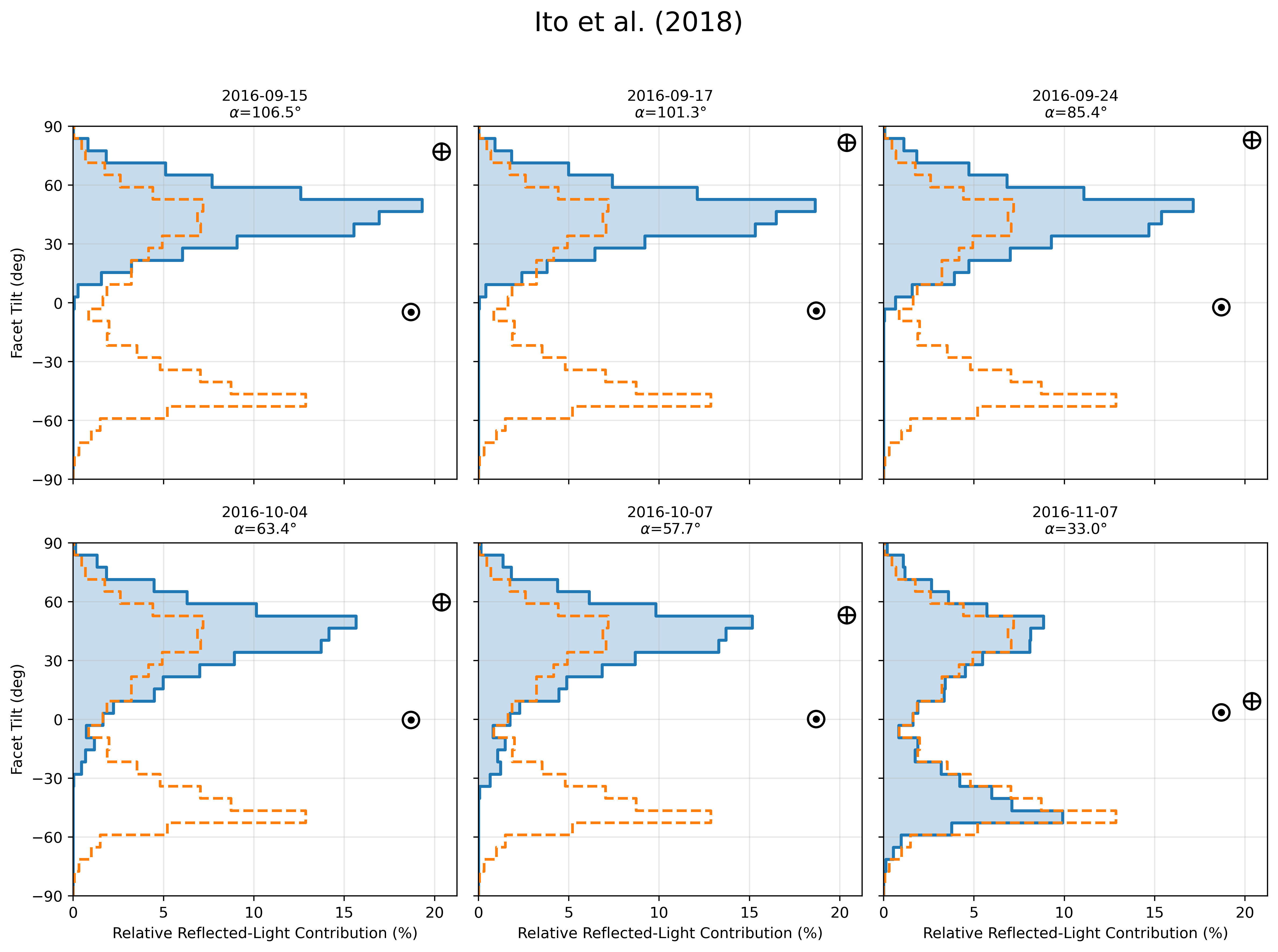}
    \caption{Modelled reflected light distributions across Phaethon's surface observed by \citet{ito2018} during 2016. Each panel corresponds to an individual observing epoch. The sub-solar and sub-observer latitudes are denoted by the symbols $\odot$ and $\oplus$, respectively. For reference, the orange dashed histogram represents the distribution for an equatorial viewing and illumination geometry ($\beta_{\odot} = \beta_{\oplus} = 0$).}
    \label{fig:ito}
\end{figure}

\begin{figure}[!ht]
    \centering
    \includegraphics[width=1\linewidth]{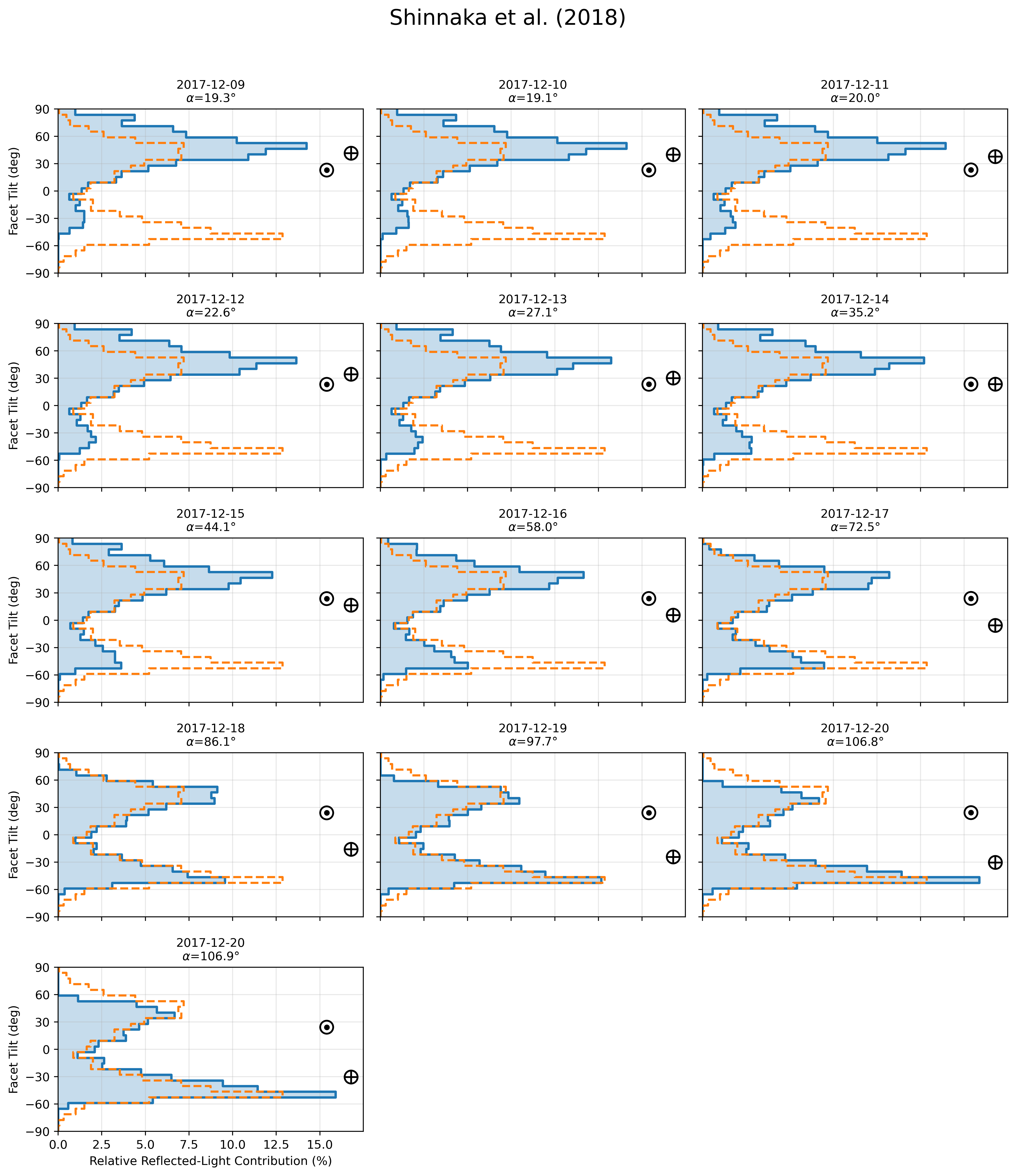}
    \caption{Same as Figure \ref{fig:ito}, but for the observations of \citet{shinnaka2018} obtained during 2017.}
    \label{fig:shinnaka}
\end{figure}

\begin{figure*}[!ht]
    \centering
    \includegraphics[width=1\linewidth]{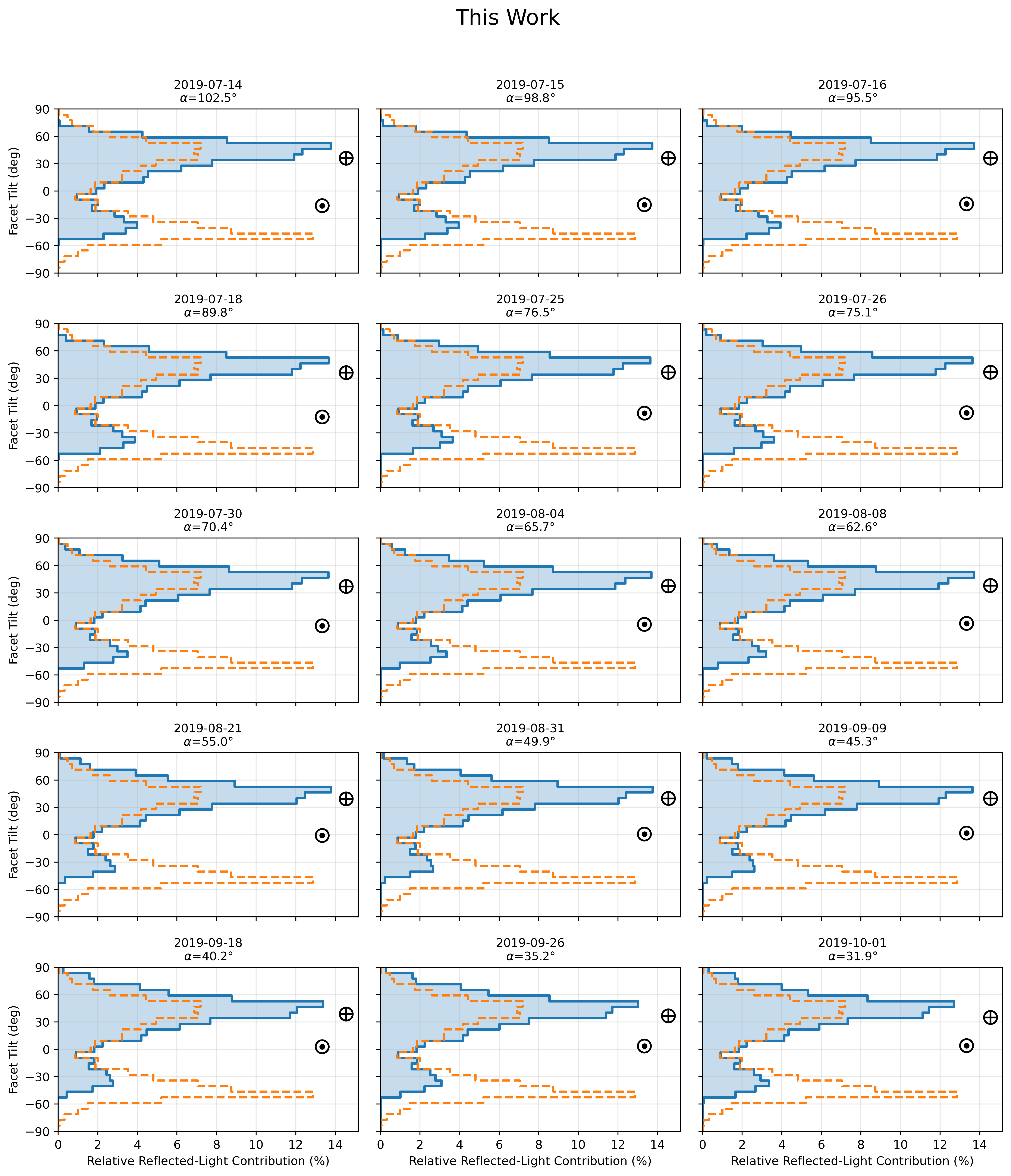}
    \caption{Same as Figure \ref{fig:ito}, but for our observations obtained during 2019.}
    \label{fig:not}
\end{figure*}

\clearpage

%TC:ignore
\section{Statistical analysis}
\label{app:stat-analysis}

To assess whether the differences in polarisation measured in 2016, 2017, and 2019 are statistically significant, we model the polarisation phase curve using the so-called trigonometric function defined in Eq.~(\ref{eq:trig}) and perform statistical analysis on the model or its residuals using two methods: the Bayesian Information Criterion (BIC) and the one-sided residual sign test. 

For analysis with BIC, we analyse three different groupings of the observations: (i) all 2016, 2017, and 2019 observations combined, (ii) 2017 and 2019 observations combined, and 2016 observations separately, and (iii) observations from each apparition separately. In the model, there are four parameters ($\alpha_0$, $b$, $c_1$, $c_2$), but from these, $b$ and $c_2$ have the largest effect on polarization values at large phase angles. With data groupings (ii) and (iii), we test models either sharing the parameters between the data groups, or having separate values. We use BIC as a criterion to select the best model between the models sharing different numbers of the parameters. BIC is defined as
\begin{equation}
    \text{BIC} = n\ln\left(\frac{\text{SSE}}{n}\right)+k\ln{n},
\end{equation}
where $n$ is the number of measurements, SSE the sum of squared residuals, and $k$ is the amount of independent parameters. Lower values of BIC correspond to a stronger support for a given model when comparing to other models using the same data. Following the BIC value, the preferred model for the case (ii) is having shared values of $\alpha_0$ and $c_1$ parameters, but separate values for $b$ and $c_2$. For the case (iii), the criterion prefers the model with shared values of $\alpha_0$ and $c_1$ among all three groups, separate value of $c_2$ for the 2016 data and the rest of the data, and separate values of $b$ for all three data groups 2016, 2017, and 2019.

Comparing the best models for cases (i)--(iii), the BIC values reveal the case (iii) to be the preferred model among all, see Table~\ref{tab:trig_pars}. The case (ii) is the second best, and using one model to fit all the data in case (i) is the least preferred. This result indicates that both the 2016 and 2019 observations are different from the 2017 observations and should be modelled (partly) independently of the 2017 observations. As the independent parameters are $b$ between all the years, and $c_2$ between 2016 and the rest, this result also hints that the difference is mainly seen with the large phase angles close to \pmax{}.

\begin{table*}[!h]
    \centering
    \caption{Best-fit parameters of the polarimetric
    phase curves calculated using Eq.~(\ref{eq:trig}) and results of the BIC analysis. $\alpha_0$ is the inversion angle; $b$, $c_1$, and $c_2$ are free parameters that shape the curve; $k$ is the combined number of independent parameters in the trigonometric fits; and $\Delta\mathrm{BIC}$ is the difference from the lowest BIC value.}
    \label{tab:trig_pars}
    \begin{tabular}{l l r r r r c c c}
        \hline
        Model & Data group & $\alpha_0$ ($^\circ$) & $b$ & $c_1$ & $c_2$ & $k$ & BIC & $\Delta\mathrm{BIC}$ \\
        \hline

        (i) All combined
        & 2016+2017+2019
        & $19.59\pm1.41$ & $25.20\pm0.89$ &  $0.923\pm0.107$ & $-1.106\pm0.075$ & 4 & $381.787$ & $245.579$ \\

        \hline

        (ii) Combined
        & 2017+2019
        & $19.92\pm0.47$ & $25.10\pm0.32$ &  $0.826\pm0.038$ & $-1.033\pm0.027$ \\

        and separate
        & 2016
        & $19.92\pm0.47$ & $25.56\pm0.66$ &  $0.826\pm0.038$ & $-1.399\pm0.059$ & 6  & $177.74$ & $41.55$ \\

        \hline

        (iii) All separate
        & 2016
        & $19.97\pm0.36$ & $25.40\pm0.54$ &  $0.760\pm0.031$ & $-1.408\pm0.048$ \\

        & 2017
        & $19.97\pm0.36$ & $24.56\pm0.26$ &  $0.760\pm0.031$ & $-1.048\pm0.022$ \\
        
        & 2019
        & $19.97\pm0.36$ & $25.53\pm0.29$ &  $0.760\pm0.031$ & $-1.048\pm0.022$ & 7 & $136.19$ & $0.00$ \\ 

        \hline
    \end{tabular}
\end{table*}

To further test the independence of the 2019 observations, we carried out a sign test for the observations to test if their systematic shift toward higher polarization relative to the jointly fitted 2017 and 2019 phase curve is statistically significant.
The sign test evaluates whether positive and negative residuals occur with equal probability. Under the null hypothesis, each residual has an equal probability of being positive or negative, and the number of positive residuals therefore follows a binomial distribution with a probability of $p=0.5$.
Of the 15 of our measurements from 2019, 14 exhibit positive residuals when fitting together the 2017 and 2019 measurements. The one-sided residual sign test gives a probability of $p=4.88\times10^{-4}$, meaning that the positive residuals are very unlikely under the null hypothesis of equal probabilities for positive and negative residuals.
Thus, the 2019 observations are systematically shifted towards higher polarization values and should be treated as an independent trend.
%TC:endignore

\end{document}